\documentclass[11pt]{article}
\usepackage{color}
\usepackage{amssymb}
\usepackage{amsmath}
\usepackage{multirow}
\usepackage{epsfig}
\usepackage{mfpic}
\usepackage{amscd}
\usepackage{feynmf}

\def\Journal#1#2#3#4{{#1} {\bf #2}, (#3) #4}

\def\etal{{\it et al.}}

\def\APH{\em Annals Phys.}

\def\APJ{\em ApJ.}
\def\APP{\em Astropart. Phys.}

\def\AST{\em Astron. J.}
\def\CAM{\em J. Comput. Appli. Math.}

\def\CPL{\em Chin. Phys. Lett.}

\def\IMD{{\em Int. J. Mod. Phys.} D}

\def\JCA{\em J. Cosmol. Astrop. Phys.}
\def\JHE{\em J. High Ener. Phys.}

\def\JPL{\em JETPhys. Lett.}

\def\LTP{\em J. Low Temp. Phys.}

\def\NPAS{{\em Nucl.Phys.Proc.Suppl.} A}
\def\NPB{{\em Nucl. Phys.} B}

\def\PLB{{\em Phys. Lett.} B}

\def\PRD{{\em Phys. Rev.} D}
\def\PRL{\em Phys. Rev. Lett.}
\def\PRV{\em Phys. Rev.}
\def\PRE{\em Phys. Rep.}

\def\be{\begin{equation}}
\def\ee{\end{equation}}
\def\bea{\begin{eqnarray}}
\def\eea{\end{eqnarray}}
\def\bes{\begin{equation*}}
\def\ees{\end{equation*}}
\def\beas{\begin{eqnarray*}}
\def\eeas{\end{eqnarray*}}

\def\tr{\text{tr}}
\def\mg{\mathsf g}
\def\cx{\mathtt X}
\def\ca{\mathtt A}
\def\um{\mathcal U}
\def\vm{\mathcal V}
\def\wm{\mathcal W}
\def\ym{\mathcal Y}
\def\zm{\mathcal Z}

\begin{document}
\begin{center}
{\Large \bf Cosmological Condensation of Scalar Fields - Making a dark energy}\\
\end{center}

\begin{center}
{\it Houri Ziaeepour\\
{
Max Planck Institut f\"ur Extraterrestrische Physik (MPE), 
Giessenbachstra$\mathbf{\beta}$e 1, 85748 Garching, Germany.\\
Email: {\tt houriziaeepour@gmail.com}}
}
\end{center}


\begin{abstract}
Our Universe is ruled by quantum mechanics and its extension Quantum Field Theory (QFT). 
However, the explanations for a number of cosmological phenomena such as inflation, dark 
energy, symmetry breakings, and phase transitions need the presence of classical scalar 
fields. Although the process of condensation of scalar fields in the lab is fairly 
well understood, the extension of results to a cosmological context is not trivial. 
Here we investigate the formation of a condensate - a classical scalar field - after 
reheating of the Universe. We assume a light quantum scalar field produced by the 
decay of a heavy particle, which for simplicity is assumed to be another scalar. 
We show that during radiation domination epoch under certain conditions, the decay of the 
heavy particle alone is sufficient for the production of a condensate. This process is very 
similar to preheating - the exponential particle production at the end of inflation. During 
matter domination epoch when the expansion of the Universe is faster, the decay alone 
can not keep the growing trend of the field and the amplitude of the condensate decreases 
rapidly, unless there is a self interaction. This issue is particularly important for dark 
energy. We show that quantum corrections of the self-interaction play a crucial role in 
this process. Notably, they induce an effective action which includes inverse power-law terms, 
and therefore can lead to a tracking behaviour even when the classical self-interaction is a 
simple power-law of order 3 or 4. This removes the necessity of having nonrenormalisable terms 
in the Lagrangian. If dark energy is the condensate of a quantum scalar field, these results 
show that its presence is deeply related to the action of quantum physics at largest 
observable scales.
\end{abstract}

\begin{fmffile}{fmfgenfydig}
\section{Introduction} \label{sec:intro}
Observations of phenomena such as superconductivity and super fluidity in condense matter 
indicates that quantum particles can collectively behave like a classical self-interacting 
scalar field. The potential energy of this interaction plays an important role in breaking 
global and/or local (gauge) symmetries which usually are followed by a phase transition. The 
same phenomena is assumed to happen at fundamental level in particle physics where usually 
a quantum scalar field, e.g. Higgs boson is responsible for dynamical mass generation. Other 
phenomena, mostly cosmological such as inflation, leptogenesis, and many of candidate models 
for dark energy are based on the existence of a classical scalar field which is usually 
related to a fundamental quantum scalar field because the physics of the Universe and its 
content in its most elementary level is quantic. 

A classical field is more than just 
classical behaviour of a large number of scalar particles. In a quantum system particles 
can be in superposition states i.e. quantum mechanically correlated to each others. 
Decoherence which is generated by interaction of each particle or field with its environment 
remove the quantum superposition and correlation between quantum states, but this does not 
mean that after decoherence of scalar particles, they behave collectively like a classical 
field. The following simple example can demonstrate this fact: \\
Consider a closed system consisting of a macroscopic amount of unstable 
massive scalar particles which decay to a pair of light scalar particles with 
a global $SU (2)$ symmetry and a very weak coupling with each other. If the 
unstable particle is a singlet of this symmetry, the remnant particles are 
entangled by their $SU (2)$ state. After a time much larger than the lifetime 
of the massive particle, the system consists of a relativistic gas of pair 
entangled particles. If a detector measures this $SU (2)$ charge without 
significant modification of their kinetic energy, the entanglement of pairs 
will break i.e. the system decoheres and becomes a relativistic gas. The 
equation of state of a relativistic ideal gas is 
$w_{rel} = P/\rho \approx 1 / 3$ where the pressure $P$ and density $\rho$ are 
defined as the expectation value of some operators acting on the Fock space 
of the system. By contrast, a classical scalar field $\varphi (x)$ is a 
$C$-number and its density ${\rho}_{\varphi}$, pressure $P_{\varphi}$, and 
kinetic energy are defined as:
\bea
&&{\rho}_{\varphi} \equiv K_{\varphi} + V (\varphi) \label{rhophi} \\
&&P_{\varphi} \equiv K_{\varphi} - V (\varphi) \label{pphi} \\
&&K_{\varphi} = \frac {1}{2}g^{\mu\nu} \partial_\mu \varphi \partial_\nu \varphi 
\label{kphi}
\eea
where $V (\varphi)$ is a potential presenting the self-interaction of the field 
$\varphi (x)$. When it is much smaller than kinetic energy $ K_{\varphi}$ , we 
obtain $P_{\varphi} \approx {\rho}_{\varphi}$, and if 
$V (\varphi) \gg K_{\varphi}$, $P_{\varphi} \approx -{\rho}_{\varphi}$. 
Therefore in general, a relativistic gas and a scalar field do not share the 
same equation of state, and the proof of decoherence in a system is not 
enough when a classical scalar field is needed to explain a physical phenomenon. 

Historically, the concept of a classical scalar field was first appeared 
in the context of scalar-tensor - Brans-Dicke - gravity theories 
(see~\cite{scalarhistory} for a historical review). In these models the 
scalar field presents dilaton, the generator of conformal symmetry. Therefore it had a purely 
geometric nature. It was only later when people tried to quantize Einstein and other gravity 
models that this field got a {\it particle} interpretation. The discovery of Higgs mechanism 
and other phenomena in condense matter in which scalar fields are present, encouraged this 
interpretation. More recently scalar field are found to be a principle ingredient in 
supersymmetric and superstring theories. In the classical limit quantum scalar fields  
are usually identified with classical fields and their differences are overlooked.

When a classical system is quantized, according to canonical quantization procedure, 
classical observables are replaced by operators acting on a Hilbert or Fock space, 
respectively for a single particle and for a multi-particle quantum system. 
The expectation values of these operators are the outcome of measurements. 
Therefore, it is natural to define the classical observable related to a 
quantum scalar field as its expectation value:
\be
\varphi (x) \equiv \langle \Psi|\Phi (x)|\Psi\rangle \label{classphi}
\ee
where $|\Psi\rangle$ is the state of the quantum system i.e. an element of the 
Fock space of the system. In analogy with particles in the ground state in quantum 
mechanics, the classical field $\varphi (x)$ is also called a {\it condensate}. In fact a 
coherent state consisting of superposition of particles in the ground state behaves like a 
classical field i.e. $\langle \Psi|\Phi (x)|\Psi\rangle \neq 0$~\cite{condwave}. This 
is an ideal and exceptional case in which the number of particles in the system is infinite. 
Nonetheless, in the cosmological context where the number of particles is very large it can 
be a good approximation. Thus, later in this work we use this state to calculate the 
evolution of a condensate in an expanding universe. Using canonical representation, 
it is easy to see that for systems with a limited number of scalar particles 
$\langle \Psi|\Phi|\Psi \rangle = 0$. But in presence of an interaction, even after 
renormalization, a finite term can survive~\cite{infrenorm} to play the role of a classical 
field (condensate) according to the definition (\ref{classphi}). In fact $\Phi$ can be 
considered to be {\it dressed} and its expectation value 
even on the vacuum can be non-zero. Equivalently, $\Phi$ can be considered 
as a free field. In this case $|\Psi\rangle$ must include infinite number of 
interacting particles. In both interpretations the presence of an interaction is a 
necessary condition for the condensation of a finite system~\cite{condint} 
(see also Ref.~\cite{noneqqft,nonequi} for a review).

Although classical scalar fields play crucial roles in the modeling of many phenomena 
particle physics and cosmology, their existence is usually considered as granted and the 
efforts are concentrated on the relevant potentials, solutions of their dynamic equations, 
and quantization of small fluctuations around the classical background fields. For instance 
in the context of inflation and reheating of the Universe, fluctuations of inflaton are 
quantized around the uniform and classical background which is responsible for the 
exponential expansion of the Universe (see e.g. Ref.~\cite{cosmoweinberg} for a review). 
Both in inflation and in ultra-cold matter the presence of a condensate is apriori 
justifiable. If the entropy of the system e.g. Universe before inflation was very small and 
inflatons were the dominant content, most of them had to be in their ground state - the zero 
mode - and therefore according to Ref.~\cite{condwave} 
behaved as a condensate (see also Appendix \ref{app:b} for a more general description of 
a condensate state). In other contexts such as in the reheating era, and in cosmological 
and lab phase transitions the entropy is not always small. Therefore, as the above example 
showed, in these cases the formation of a condensate from a quantum scalar field is not a 
trivial process and the necessary conditions for the existence of such coherent behaviour 
must be investigated.

In quintessence models a classical scalar field is the basic content of the model and its 
energy density is interpreted as dark energy. Although in the framework of popular particle 
physics models such as supersymmetry, supergravity, and string theory many efforts have 
been concentrated on finding candidate scalar fields to play the role of 
quintessence~\cite{quincand}, little work has been devoted to understand what are the 
necessary conditions for a quantum scalar field to condense in a manner which satisfies the very 
special characteristics of a quintessence field. For instance, such a 
condensate must initially have a very small density, much smaller than other content of the 
Universe (smallness problem). Present observations show that dark energy behaves very 
similar to a cosmological constant i.e. with the expansion of the Universe its energy 
density does not change or varies very slowly. Such a behaviour is not 
trivial. In the classical quintessence models usually the potential of the model is 
{\it designed} such that a tracking solution do exist. Potentials with 
such property are usually non-normalizable. Moreover, they don't directly correspond 
to potentials (or kinetic terms) expected from fundamental theories such as 
supersymmetry, supergravity or string theory. Therefore one has to relate them ad-hocly to 
some sort of low energy effective model of a fundamental theory.

The purpose of the present work is to fill the gap between quantum processes producing 
various species of particles/fields in the early Universe - presumably during and after 
reheating - and their classical component as defined in (\ref{classphi}). In another word, 
we want to see how the microscopic properties of matter is related to macro-physics and vis 
versa. We are particularly interested in condensation at very large scales, relevant to 
dark energy models. For other phenomena such as baryo- and lepto-genesis and Higgs mechanism, 
if the energy scale was much larger than Hubble constant of the epoch, the process can be 
studied locally. This is not the case for dark energy which seems to be uniform at largest 
observable scales, and therefore the expansion of the Universe could play important role in 
its evolution.

As the quantum physics of the epoch just after reheating is not well known, we consider 
a simple toy model with a light scalar as quintessence field in interaction with 
two heavy scalar fields. Our aim is to study the evolution of the 
classical component - the condensate - of the quintessence field. Between many possible 
types of quantum scalar field and interaction models, we concentrate on a class of models 
in which the heaviest of three particle decays to other fields. The motivation for such a 
model is the results obtained from the study of the effect of a decaying dark matter 
on the equation of states of the Universe~\cite{houriustate}. It has been 
shown that a FLRW cosmology with a decaying dark matter and a cosmological 
constant behaves similar to a cosmology with a stable dark matter and a dark energy 
with $w = P/\rho \lesssim -1$. This is effectively what is concluded at least from some of 
present supernovae observations~\cite{snobs}. More recently, the same effect has been 
proved to exist for the general case of interaction between dark matter and dark 
energy~\cite{quindmint}. 
It has been also shown~\cite{houridmquin} that if a decaying dark matter has a small 
branching factor to a light scalar field, the observed density and equation of state of 
dark energy can be explained without extreme fine-tuning of the potential or coupling constants. 
In other word, such a model solves both smallness and coincidence problems of dark energy. 
The present work should complete this investigation by studying the formation and evolution 
of classical component from a quantum point of view. More generally, it is believed that all the 
particles are produced directly or indirectly from the decay of inflaton or curvaton (in 
curvaton inflation models) oscillation. Quintessence field is not an exception and 
irrespective of the details of its physics, it has to be produced from the decay of the 
inflaton or another field. Although the toy model considered here basically assumes a long 
life heavy particle, in each step of calculation we also mention the differences in the results 
if the life time of the decaying particle is short. The main difference between these cases 
is the time duration in which the production of quintessence scalar by the decay is significant.

In Sec. \ref{sec:decay} we describe the Lagrangian of a decaying dark matter model and 
evolution equation of the condensate. We consider three decay modes for the heavy particle 
and use the closed time path integral method to calculate the contribution of interactions 
in the condensate evolution. The same methodology has been used for studying inflation 
models~\cite{infquant}, late-time warm inflation~\cite{warminf}, the effects of 
renormalization and initial conditions on the physics of inflation~\cite{infrenorm}, 
baryogenesis~\cite{baryo}, and coarse-grained formulation of decoherence~\cite{coarsegrain} 
(see also~\cite{nonequi} and references therein). In Sec.\ref{sec:solevol} we solve field 
equations and discuss their boundary conditions. In Sec. \ref{sec:classevol} we obtain an 
analytical expression for the asymptotic behaviour of the condensate and discuss the 
importance of the quantum corrections. We summarize the results in Sec. \ref{sec:conclu}. 
In Appendix \ref{app:a} we obtain non-vacuum Green's functions in presence of a condensate. 
In Appendix \ref{app:b} we generalize the description of a condensate to a system in which 
not all the particles are in the ground state. Appendix \ref{app:c} presents the solution of  
evolution equations in matter dominated era. Finally, in Appendix \ref{app:d} propagators in a 
fluctuating background metric are determined.

\section{Decay in an Expanding Universe} \label{sec:decay}
We consider a simple decay mode for a heavy particle $X$ to a remnant that includes only 2 
types of particles: a light scalar $\Phi$ - light with respect to decaying particle $X$ - and 
another field $A$ of an arbitrary type. In fact, in a realistic particle physics model, most 
probably $A$ will not be a final stable state and decays/fragments to other particles. 
Therefore it should be considered as an intermediate state or a collective notation for other 
fields. In the simplest case studied here all the particles are assumed to be scalar. 
Extension to cases where the decaying particle $X$ and one of the remnants are spinors is 
straightforward. The quintessence field $\Phi$ however, must be a scalar. We do not consider 
the condensation of vector fields here. In the extreme density of the Universe after 
reheating, apriori the formation of Cooper-pair composite scalars from fermions is also 
possible. This process needs a relatively strong interaction between fermions and can arise 
in local phenomena such as Higgs mechanism and leptogenesis which 
happen at high energies (short distances)~\cite{composhiggs}, but not for dark energy which 
must have a very weak interaction and acts at cosmological scales. 

We consider the following decay modes:
\bea
&&(a) \quad \quad \quad \quad \quad \quad \quad \quad \quad
(b) \quad \quad \quad \quad \quad \quad \quad \quad \quad \quad (c) \nonumber \\
&&\begin{fmfgraph*}(30,30)
\fmfleft{i1}
\fmfright{o2,o3}
\fmf{dbl_plain_arrow,label=$X$}{i1,v1}
\fmf{fermion,label=$A$}{v1,o2}
\fmf{plain_arrow,label=$\Phi$}{v1,o3}
\fmfdot{v1}
\end{fmfgraph*} \quad \quad \quad
\begin{fmfgraph*}(30,30)
\fmfleft{i1}
\fmfright{o2,o3,o4}
\fmf{dbl_plain_arrow,label=$X$}{i1,v1}
\fmf{fermion,label=$A$}{v1,o2}
\fmf{plain}{v1,v2}
\fmf{plain_arrow,label=$\Phi$}{v2,o3}
\fmf{fermion,label=$A$}{v1,o4}
\fmfdot{v1}
\end{fmfgraph*} \quad \quad \quad
\begin{fmfgraph*}(30,30)
\fmfleft{i1}
\fmfright{o2,o3,o4}
\fmf{dbl_plain_arrow,label=$X$}{i1,v1}
\fmf{plain_arrow,label=$\Phi$}{v1,o2}
\fmf{plain}{v1,v2}
\fmf{fermion,label=$A$}{v2,o3}
\fmf{plain_arrow,label=$\Phi$}{v1,o4}
\fmfdot{v1}
\end{fmfgraph*} \label{decaymode}
\eea
Diagram (\ref{decaymode}-a) is the simplest decay/interaction mode. Diagram (\ref{decaymode}-b) 
is a prototype decay mode when $X$ and $\Phi$ share a conserved quantum number or $A$ and 
$\bar{A}$ (here $A=\bar{A}$ is considered) has a conserved quantum number. For instance, one of 
the favorite candidates for $X$ is a sneutrino decaying to a much lighter scalar field 
(e.g. another sneutrino) carrying the same leptonic number~\cite{rnu}~\cite{rnu1}. With 
seesaw mechanism in the superpartner sector (or even without it~\cite{rnu1}) if SUSY breaking 
scale is lower than seesaw scale, a mass split between right and left neutrinos and 
sneutrinos will occur. As the right-hand neutrino super-field is assumed to be a singlet of 
the GUT gauge symmetry, it has only Yukawa-type of interaction. In such a setup $X$ can be a 
heavy right sneutrino decaying to a light sneutrino with the same leptonic number and a pair 
of Higgs or Higgsino~\cite{rnudm}. In place of assuming two $A$ particles in the final state 
we could consider them as being different $A$ and $\bar{A}$, its anti-particle. But this adds 
a bit more complexity to the model and does not change its general behaviour. For this reason 
we simply consider the same field. Diagram (\ref{decaymode}-c) is representative of a 
case where $X$ and $A$ can be fermions (although we do not consider this case 
here) and $\Phi$ can be complex and carries a conserved charge~\cite{snuaxion} 
(again for simplicity we do not consider this case here either).

The corresponding Lagrangians of these effective interactions are the 
followings:
\bea
{\mathcal L}_{\Phi} &=& \int d^4 x \sqrt{-g} \biggl [\frac{1}{2} g^{\mu\nu}
{\partial}_{\mu}\Phi {\partial}_{\mu}\Phi - \frac{1}{2}m_{\Phi}^2 {\Phi}^2 - 
\frac{\lambda}{n}{\Phi}^n \biggr ] \label{lagrangphi} \\
{\mathcal L}_{X} &=& \int d^4 x \sqrt{-g} \biggl [\frac{1}{2} g^{\mu\nu}
{\partial}_{\mu}X {\partial}_{\mu}X - \frac{1}{2}m_X^2 X^2 \biggr ] 
\label{lagrangx}\\
{\mathcal L}_{A} &=& \int d^4 x \sqrt{-g} \biggl [\frac{1}{2} g^{\mu\nu}
{\partial}_{\mu}A {\partial}_{\mu}A - \frac{1}{2}m_A^2 A^2 - 
\frac{\lambda'}{n'}A^{n'} \biggr ] \label{lagranga} \\
{\mathcal L}_{int} &=& \int d^4 x \sqrt{-g} \begin{cases} 
\mg \Phi X A, & \text{For (\ref{decaymode})-a} \\
\mg \Phi X A^2, & \text{For (\ref{decaymode})-b} \\ 
\mg {\Phi}^2 XA, & \text{For (\ref{decaymode})-b} \end{cases}  
\label{lagrangint}
\eea
In addition to the interaction between $X$, $\Phi$, and $A$ we have assumed a 
power-law self-interaction for $\Phi$ and $A$. If $A$ is a collective 
notation for other fields in more realistic models, its self-interaction 
corresponds to the interaction between these unspecified fields. Again for 
the sake of simplicity in the rest of this work we consider $\lambda' = 0$. 
The unstable particle $X$ is assumed to have no self-interaction. Note that 
the same Lagrangian can be considered to present the interaction between 
these fields. Therefore with little modification, the results of this work 
become applicable to the case of interaction between dark matter and dark 
energy.

Although the model presented here is quite general, we are primarily interested on the 
physics of dark energy. In this context, the heavy particle $X$ is a candidate for the dark 
matter, and $\Phi$ is the quintessence field, and the Lagrangian (\ref{lagrangint}) 
presents the interaction between these fields. It is 
necessary that $X$ and $\Phi$ have a very weak interaction with each other and with the rest 
of the Universe presented by $A$. Therefore couplings $\lambda$ and $\mg$ must be very small. 

In a realistic particle physics model, renormalization as well as non-perturbative effects 
can lead to complicated potentials for the scalar fields. An example relevant to dark energy 
is a pseudo-Nambu-Goldston boson field. Its potential is assumed to have a shift 
symmetry~\cite{quinpngb}. This class of models are interesting for the fact that the mass of the 
quintessence field does not receive quantum corrections and can be very small. Moreover, 
they can be easily implemented in SUSY theories along with right-neutrinos and 
sneutrinos, as candidate for $X$~\cite{pngsusy}. The power-law potential considered here 
can be interpreted as the dominant or one of the terms in the polynomial expansion of the 
potential. In addition, we Will see in Sec.\ref{sec:solevol} that only the few lowest order in 
the expansion play a significant role in the late time behaviour of the condensate. 

The general aspects of the analysis presented here do not depend on the details of the 
particle. Here our aim is an analytical 
investigation of the evolution of the condensate to see whether it is possible at all to have 
a quantum condensate at large scales. To achieve this goal we had to apply many approximations 
and simplifications. A more precise solution needs numerical analysis and we leave it to a 
future work.

\subsection{Decomposition} \label{sec:decomp}
We decompose $\Phi (x)$ to a classical (condensate) and quantum components:
\be
\Phi (x) = \varphi (x) I + \phi (x) \quad \quad \langle \Phi \rangle \equiv 
\langle \Psi|\Phi|\Psi \rangle = \varphi (x) \quad \quad  \langle \phi \rangle 
\equiv \langle \Psi|\phi|\Psi \rangle = 0 \label{decomphi} 
\ee
where $I$ is the unit operator. Note that in (\ref{decomphi}) both classical 
and quantum components depend on the spacetime $x$. In studying inflation 
it is usually assumed that the pre-inflation Universe was homogeneous or the 
very fast expansion of the Universe had washed out all the inhomogeneities 
and the condensed component of inflaton became homogeneous. Here we are 
studying the evolution of a scalar field after inflation when the distribution 
of the unstable $X$ particles can have non-negligible inhomogeneities, 
specially if the decay is slow and perturbations have time to grow. Thus, we don't 
assume a homogeneous Universe, but anisotropies are assumed to be small. Moreover,
for solving dynamic equations, in some situations we have to ignore anisotropies 
all together, otherwise the problem would not be tractable analytically.

We assume $\langle X \rangle = 0$ and $\langle A \rangle = 0$. Justification 
for these assumptions is the large mass and small coupling of $X$ and $A$ 
which should 
reduce their number and their quantum correlation. In other words, when mass 
is large, the minimum of the effective potential for the classical component 
is pushed to zero (see (\ref{dyneffa}-\ref{dyneffc}) below). We find a 
quantitative justification for negligible condensation of massive fields in 
Sec. \ref{sec:classevol}. As $X$ and $A$ have a very weak 
interaction with $\phi$ and the condensate, their evolution can be studied 
semi-classically by simply considering the decay and interaction cross-sections. 
Such a study has been already performed in~\cite{houriustate, houridmquin}. 
Therefore, here we concentrate on the evolution of the condensate $\varphi (x)$ 
and if necessary, we use some of the results from the works mentioned above.

The Lagrangian of $\Phi$ is decomposed to:
\bea
{\mathcal L}_{\Phi} & = & {\mathcal L}_{\varphi} + {\mathcal L}_{\phi} + 
{\mathcal L}_{int} \label{lagrangeff} \\
{\mathcal L}_{int} & = &\int d^4 x \sqrt{-g} \biggl [-\frac{\phi}{2 \sqrt{-g}}
\biggl ({\partial}_{\mu}(\sqrt{-g} g^{\mu\nu} {\partial}_{\nu}\varphi) + 
{\partial}_{\nu}(\sqrt{-g} g^{\mu\nu} {\partial}_{\nu}\varphi) \biggr ) - 
m_{\Phi}^2\varphi\phi - \frac{\lambda}{n}\sum_{i=0}^{n-1} 
\binom{n}{i}{\varphi}^i{\phi}^{n-i} \biggr ] + \nonumber \\
& & \int d^4 x \sqrt{-g} \begin{cases} \mg\varphi XA +
\mg \phi XA & \quad \quad \text{For (\ref{decaymode})-a} \\
\mg\varphi XA^2 + \mg \phi XA^2 & \quad \quad \text{For (\ref{decaymode})-b} \\
\mg{\varphi}^2 XA + 2\mg\varphi \phi XA + \mg {\phi}^2 XA & \quad \quad 
\text{For (\ref{decaymode})-c}
\end{cases} \label {lagrangeint}
\eea
The first two terms in ${\mathcal L}_{int}$ are obtained after integrating out two total 
derivative terms. Lagrangians ${\mathcal L}_{\varphi}$ and ${\mathcal L}_{\phi}$ are the 
same as (\ref{lagrangphi}) with $\Phi \rightarrow \varphi$ and $\Phi \rightarrow \phi$ 
respectively. The self-interaction terms in (\ref{lagrangeint}) include a term proportional 
to $\phi^2$ that contributes to the mass of quantum component. Therefore, when we use the 
free Lagrangian of the quantum component to compute quantum corrections, this term is 
considered to belong to ${\mathcal L}_{\phi}$. 

The evolution equation for the condensate (classical component) $\varphi$ can be obtained 
from Lagrangian $L_{\Phi}$ by variation method. we must also take the expectation value 
of operators on the state $|\Psi\rangle$:
\bea
\frac{1}{\sqrt{-g}}{\partial}_{\mu}(\sqrt{-g} g^{\mu\nu}{\partial}_{\nu}
\varphi) + m_{\Phi}^2 \varphi + \frac{\lambda}{n}\sum_{i=0}^{n-1} (i+1)
\binom{n}{i+1}{\varphi}^i\langle{\phi}^{n-i-1}\rangle - \mg \langle 
XA\rangle = 0 && \nonumber \\
\hspace {12cm} \text{For (\ref{decaymode})-a} && \label {dyneffa} \\ 
\frac{1}{\sqrt{-g}}{\partial}_{\mu}(\sqrt{-g} g^{\mu\nu}{\partial}_{\nu}
\varphi) + m_{\Phi}^2 \varphi + \frac{\lambda}{n}\sum_{i=0}^{n-1} (i+1)
\binom{n}{i+1}{\varphi}^i\langle{\phi}^{n-i-1}\rangle - \mg \langle 
XA^2\rangle = 0 && \nonumber \\
\hspace {12cm} \text{For (\ref{decaymode})-b} && \label {dyneffb} \\ 
\frac{1}{\sqrt{-g}}{\partial}_{\mu}(\sqrt{-g} g^{\mu\nu}{\partial}_{\nu}
\varphi) + m_{\Phi}^2 \varphi + \frac{\lambda}{n}\sum_{i=0}^{n-1} 
(i+1)\binom{n}{i+1}{\varphi}^{i}\langle{\phi}^{n-i-1}\rangle - 2\mg\varphi 
\langle XA\rangle - 2\mg \langle \phi XA\rangle = 0 && \nonumber \\
\hspace {12cm} \text{For (\ref{decaymode})-c} && \label {dyneffc}
\eea
Note that in (\ref{lagrangeint}) non-local interactions, i.e. terms containing derivatives 
of $\varphi$ do not contribute in the evolution of $\varphi$ because they are all 
proportional to $\phi$. After taking expectation value of the operators they cancel out 
because $\langle \phi \rangle = 0$ by definition. The expectation values depend 
on the quantum state of the system $|\Psi\rangle$ which presents the state all the particles 
in the system. From the structure of Lagrangians (\ref{lagrangeff}) and (\ref{lagrangeint}) 
it is clear that the mass of quantum component $\phi$ and thereby its evolution depends on 
$\varphi$. Moreover, through the interaction of $\Phi$ with $X$ and $A$ the evolutions of all 
the constituents of this model are coupled. See Appendix \ref{app:a} for more details.

For $n \geqslant 2$ the expectation values $\langle \phi^{(n-i-1)} \rangle$ 
modify the mass and self-coupling of $\varphi$. Another important observation is that 
in general the form of the potential for the classical field $\varphi$ is not the same as 
the potential in the original Lagrangian, although they have the same order. Therefore, 
the usual practice in the literature of using the same potential for both quantum and 
classical component is only an approximation.

For models (a) and (b) the expectation value of interaction in (\ref{dyneffa}) and 
(\ref{dyneffb}) contains only $X$ and $A$. But when these terms are expanded, see equations 
(\ref{avalxa}) and (\ref{valxaa}) below, they depend also on $\varphi$. We will see later 
that these terms play the role of a feedback between production and evolution of the 
classical field. In particular, they prevent a complete decay of the condensate with the 
accelerating expansion of the universe. This effect is similar to what was found in the 
classical treatment of the same models in Ref.~\cite{houridmquin}. Interaction (c) is more 
complex and various evolution histories for $\varphi$ are possible. They depend on the value 
and sign of $\mg$ the coupling of $\Phi$ to $X$, self-coupling $\lambda$, and the order of 
self-interaction potential $n$. For instance, the mass can become imaginary (tachyonic) 
even without self-interaction and lead to a symmetry breaking. Tachyonic scalar fields 
have been suggested as quintessence field specially in the framework of models with 
$w < -1$~\cite{tachyonquin}. We discus the difference between these decay modes at each 
step of calculation. 

\subsection{Expectation values} \label{sec:expval}
We use Schwinger closed time path formalism also called in-in to calculate expectation 
values. Recent reviews of this formalism are available~\cite{ctprev,nonequi} and here we 
only present the results. Zero-order (tree) diagrams for the expectation values in 
(\ref{dyneffa}-\ref{dyneffc}) are shown in (\ref{xaadiag}), (\ref{xadiag}) and 
(\ref{xaphidiag}). The next relevant diagrams are of order $g^3$ and for dark energy models 
are expected to be negligibly small. Evidently the decomposition of $\Phi$ also affects 
the renormalization of the model. This issue has been already studied~\cite{infrenorm} and 
we do not consider it here. One example of higher order diagrams is shown in 
(\ref{xathreediag}). These types of graphs are specially important for studying 
renormalization in the context of a realistic particle physics model. Thus, for the 
phenomenological models considered here we ignore them.
\be
\langle XA \rangle_a = \quad  
\parbox{80mm}{\begin{fmfgraph*}(30,20)
\fmfleft{i1}
\fmfright{o2,o3}
\fmf{dashes,label=$\varphi$}{i1,v1}
\fmf{plain,label=$A$}{v1,o2}
\fmf{dbl_plain,label=$X$}{v1,o3}
\fmfdot{v1}
\end{fmfgraph*} \hspace{0.5cm} + \hspace{0.5cm}
\begin{fmfgraph*}(30,20) 
\fmfleft{i1}
\fmfright{o2,o3}
\fmf{dashes,label=$\varphi$}{i1,v1}
\fmf{dbl_plain,label=$X$,tension=1/2}{v1,v2}
\fmf{plain,label=$\phi$,tension=1/2}{v2,v3}
\fmf{plain,label=$A$,tension=1/2}{v1,v3}
\fmf{plain,label=$A$}{v2,o2}
\fmf{dbl_plain,label=$X$}{v3,o3}
\fmfdot{v1}
\fmfdot{v2}
\fmfdot{v3}
\end{fmfgraph*}} \hspace{1cm} + \ldots \label {xathreediag}
\ee
\bea
\langle XA^2\rangle = \quad  
\parbox{40mm}{\begin{fmfgraph*}(30,20)
\fmfleft{i1}
\fmfright{o2,o3,o4}
\fmf{dashes,label=$\varphi$}{i1,v1}
\fmf{plain,label=$A$}{v1,o2}
\fmf{dbl_plain}{v1,v2}
\fmf{dbl_plain,label=$X$}{v2,o3}
\fmf{plain,label=$A$}{v1,o4}
\fmfdot{v1}
\end{fmfgraph*}} \hspace{1cm} & + & \ldots \label {xaadiag} \\
\langle XA\rangle_c = \quad  
\parbox{40mm}{\begin{fmfgraph*}(40,20)
\fmfleft{i1,i2}
\fmfright{o3,o4}
\fmf{dashes,label=$\varphi$}{i1,v1}
\fmf{dashes,label=$\varphi$}{i2,v1}
\fmf{plain,label=$A$}{v1,o3}
\fmf{dbl_plain,label=$X$}{v1,o4}
\fmfdot{v1}
\end{fmfgraph*}} \hspace{1cm} & + & \ldots  \label {xadiag} \\
\langle \phi XA\rangle = \quad  
\parbox{40mm}{
\begin{fmfgraph*}(30,20)
\fmfleft{i1}
\fmfright{o2,o3,o4}
\fmf{dashes,label=$\varphi$}{i1,v1}
\fmf{dbl_plain,label=$X$}{v1,o2}
\fmf{plain}{v1,v2}
\fmf{plain,label=$\phi$}{v2,o3}
\fmf{plain,label=$A$}{v1,o4}
\fmfdot{v1}
\end{fmfgraph*}} \hspace{1cm} & + & \ldots \label {xaphidiag} \\
\langle{\phi}^{i}\rangle = \quad
\parbox{40mm}{
\begin{fmfgraph*}(30,20)
\fmfleft{i1,i2,i3,i4}
\fmfright{o5,o6,o7}
\fmf{dashes,label=$\varphi$}{i1,v1}
\fmf{dashes,label=$\varphi$}{i2,v1}
\fmf{dashes,label=$\varphi$}{i3,v1}
\fmf{dashes,label=$\varphi$}{i4,v1}
\fmf{plain,label=$\phi$,tension=1.5}{v1,o5}
\fmf{plain,label=$\phi$,tension=1.5}{v1,o6}
\fmf{plain,label=$\phi$,tension=1.5}{v1,o7}
\fmfdot{v1}
\end{fmfgraph*}} \hspace{1cm} & + & \ldots \label {phiphidiag}
\eea
The index $a$ and $c$ in (\ref{xathreediag}) and (\ref{xadiag}) refer to the corresponding 
interaction model respectively. The graph (\ref{phiphidiag}) is an example of 
self-interaction terms in (\ref{dyneffa}-\ref{dyneffc}) with $n$ external lines where 
$i \geqslant 1$ of them are of type $\phi$ and the rest of type $\varphi$. The dash lines 
present the classical component $\varphi$. When $i = 1$ there is an additional interaction 
involving derivative of the classical field.
\be
\langle \phi \rangle = \quad \parbox{40mm}{
\begin{fmfgraph*}(30,20)
\fmfleft{i1}
\fmfright{o1}
\fmfblob{.1w}{v1}
\fmf{plain}{v1,i1}
\fmf{dashes,label=$\frac{1}{\sqrt{-g}}{\partial}_{\mu}(\sqrt{-g} g^{\mu\nu}{\partial}_{\nu}) + m_{\Phi}^2$}{o1,v1}
\end{fmfgraph*}} \hspace{1cm} + \quad
\parbox{40mm}{
\begin{fmfgraph*}(30,20)
\fmfleft{i1}
\fmfright{o1,o2,o3}
\fmf{plain,label=$\phi$,tension=1.5}{v1,i1}
\fmf{dashes,label=$\varphi$}{o1,v1}
\fmf{dashes,label=$\varphi$}{o2,v1}
\fmf{dashes,label=$\varphi$}{o3,v1}
\fmfdot{v1}
\end{fmfgraph*}} \label {phiphivardiag}
\ee
At lowest order the sum of these graphs is null because they correspond to the dynamic 
equation of $\varphi$ see (\ref{dyneffa}-\ref{dyneffc}). This is consistent with the 
decomposition (\ref{decomphi}).

The corresponding expectation values at zero order are:
\bea
\langle XA\rangle_a & = & -i\mg \int \sqrt{-g} d^4y \varphi (y)
\biggl [G_A^> (x,y) G_X^> (x,y) - G_A^< (x,y) G_X^< (x,y)\biggr ] 
\label {avalxa} \\
\langle XA^2\rangle & = & -i\mg \int \sqrt{-g} d^4y \varphi (y)
\biggl [G_A^> (x,y) G_A^> (x,y) G_X^> (x,y) - G_A^< (x,y) 
G_A^< (x,y) G_X^< (x,y)\biggr ] \label {valxaa}\\
\langle XA\rangle_c & = & -i\mg \int \sqrt{-g} d^4y {\varphi}^2 (y)
\biggl [G_A^> (x,y) G_X^> (x,y) - G_A^< (x,y) G_X^< (x,y)\biggr ] 
\label {valxa} \\
\langle \phi XA\rangle & = & -i\mg \int \sqrt{-g} d^4y \varphi (y)
\biggl [G_{\phi}^> (x,y) G_A^> (x,y) G_X^> (x,y) - 
G_{\phi}^< (x,y) G_A^< (x,y) G_X^< (x,y)\biggr ] \label {valxaphi} \\
\langle \phi^i\rangle & = & -i\lambda \int \sqrt{-g} d^4y \varphi^{n-i} (y)
\biggl [[G_{\phi}^> (x,y)]^i - [G_{\phi}^< (x,y)]^i \biggr ] \label {valphiphi}
\eea
Advanced and retarded propagators $G^>$ and $G^<$ are defined as:
\bea
G^> (x,y) &\equiv& -i \langle \psi (x) {\psi}^{\dagger}(y) \rangle = 
-i \tr (\psi (x) {\psi}^{\dagger}(y)\rho) \label{proggt}\\
G^< (x,y) &\equiv& \mp i \langle {\psi}^{\dagger} (y) \psi (x) \rangle = 
\mp i \tr ({\psi}^{\dagger} (y) \psi (x)\rho)
\label{progls}
\eea
where $\psi (x)$ presents one of $\phi$, $X$ or $A$ fields and 
$\rho = |\Psi\rangle\langle\Psi|$ is the density (projection) operator for the 
state $|\Psi\rangle$. The upper and lower signs in (\ref{progls}) are 
respectively for bosons and fermions. Definitions (\ref{proggt}) and 
(\ref{progls}) correspond to the general case of a complex field. Here we only 
consider real fields and therefore $\psi (x) = {\psi}^{\dagger}(x)$. Feynman 
propagators are related to $G^> (x,y)$ and $G^< (x,y)$:
\bea
G_F (x,y) &\equiv& -i \langle T\psi (x) {\psi}^{\dagger}(y) \rangle = 
G^> (x,y) \Theta (x^0-y^0) + G^< (x,y) \Theta (y^0-x^0)\label{progf}\\
\bar{G}_F (x,y) &\equiv& -i \langle \bar{T}\psi (x) {\psi}^{\dagger}(y)
\rangle = G^> (x,y) \Theta (y^0-x^0) + G^< (x,y) \Theta (x^0-y^0)
\label{progfbar}
\eea
The next step is the calculation of propagators.

\subsection {Propagators and the evolution of quantum components} 
\label{sec:prog}
Feynman propagators $G^i_F (x,y), ~i = \phi,~X,~A~$ can be determined using 
field equations from Lagrangians (\ref{lagrangphi})-(\ref{lagrangint}). The 
free equations of motion lead to the following equations for the propagators: 
\bea
&&\frac{1}{\sqrt{-g}}{\partial}_{\mu}(\sqrt{-g} g^{\mu\nu}{\partial}_{\nu}
G^{\phi}_F (x-y)) + (m_{\Phi}^2 + (n-1) \lambda {\varphi}^{n-2}) 
G^{\phi}_F (x-y) = -i \frac{{\delta}^4 (x-y)}{\sqrt{-g}} \nonumber \\
&& \label {propagphi} \\
&&\frac{1}{\sqrt{-g}}{\partial}_{\mu}(\sqrt{-g} g^{\mu\nu}{\partial}_{\nu}G^i_F 
(x-y)) + m_i^2 G^i_F (x-y) = -i\frac{{\delta}^4 (x-y)}{\sqrt{-g}} \quad , \quad
i = X, A \label {propagx}
\eea
The free propagator of $\phi$ is independent of the type of interaction between $X$, $A$, and 
$\Phi$, and therefore equation (\ref{propagphi}) is valid for all the interaction models in 
(\ref{decaymode}). Note that $G^{\phi}_F (x-y)$ is coupled to the condensate field 
$\varphi$ even at classical level. On the other hand, evolution equations 
(\ref{dyneffa})-(\ref{dyneffc}) depend on the interaction between quantum fields $\Phi$, 
$X$ and $A$. This means that all the components of the model are coupled even at lowest 
order. The coupling between quantum component $\phi$ and the 
classical component $\varphi$ is the origin of the back-reaction of the condensate formation 
on the quantum fields. Its strength depends on the mass, order, and strength of the 
self-interaction. Assuming that initially $\Phi$ particles are produced only through the decay 
of $X$, the initial value of $\varphi = 0$. 
Therefore initially the coupling between $\phi$ and $\varphi$ was very small. With the growth 
of the $\varphi$ amplitude, the effective mass of $\phi$ particles becomes larger than their 
bare mass. On the other hand, this affects the growth of the condensate because due to an 
energy barrier $\phi$ particles will not be able to join the condensate anymore. Therefore, 
there is a negative feedback that prevents an explosive formation of the condensate.

Assuming a quick, roughly immediate decoherence of $\phi$ and other species
\footnote{This assumption 
seems inconsistent. In one hand we want that $\Phi$ has a very weak interaction with itself 
and with other particles. On the other hand we simplify the problem by assuming that they 
decohere quickly. We must also remind that if $X$ particles that produce $\Phi$ are decohered, 
so do $\Phi$ particles.}, if their interaction is weak, we do not need to consider a complete 
quantum treatment of this model\footnote{For a complete treatment one has to use the techniques 
of non-equilibrium quantum field theory and Kadanoff-Baym equations (see for 
instance~\cite{noneqqft,nonequi}) which lead to quantum Boltzmann equations 
(see~\cite{qmboltz} and references therein.)}, 
and therefore the quantum state of free particles $|\Psi_f\rangle$, including $\phi$, can be 
approximated by direct multiplication of single particle states:
\be
|\Psi_f\rangle = \sum_{p_j} \bigotimes_{i,j} f^i (x, \{p_j\})|p^i_j\rangle  \label{psif}
\ee
The indices $i$ and $j$ present the species type and particle number respectively, and 
$\{p_j\}$ the momentum of all states. In the 
context of dark energy models single particle description is a good approximation because 
the interaction of $\Phi$ with itself and with other particles must be very small. This can 
be not generalized to process in which fields can have strong couplings, such as supersymmetric 
models before SUSY breaking and electroweak in the early Universe, QCD, etc. In this case a 
complete N-particle evolution of the unfactorizable wave functions must be 
considered~\cite{noneqqft,nonequi}.

In the Introduction we argued that the expectation value of $\Phi$ on the state 
$|\Psi_f\rangle$ is null, $\langle \Psi_f|\Phi|\Psi_f\rangle = 0$. The complete state of 
the Universe $|\Psi\rangle$ depends on both condensate and free particles i.e. 
$\Psi \equiv |\Psi_f, \varphi \rangle$. When the coupling between these components is small 
$\Psi$ can be factorized:
\be
|\Psi \rangle \equiv |\Psi_f, \varphi \rangle \approx |\Psi_f\rangle \otimes 
|\varphi \rangle \label{totpsi}
\ee
A special description for $|\varphi \rangle$ is given in Appendix \ref{app:a} and a more 
general one in Appendix \ref{app:b}. The presence of $\varphi$ in $\Psi_f$ in the definition 
of $|\Psi \rangle$ reflects the fact that because of their interaction we can not completely 
separate these subsystems. The amplitudes of single-particle states $f^i$ - the one-particle 
distribution functions - in (\ref{psif}) are considered to depend on the spacetime coordinates 
to reflect the process of squeezing and decoherence of the wave functions. In fact in this 
setup, the only difference between squeezed (classical) particles of the same species is 
their place with respect to the random fluctuations of the background. More precisely 
quantization of inflaton and other fields induces a quantized metric fluctuation - 
corresponding to a semi-classical treatment of gravity - because the metric is related to 
matter through Einstein equations (see e.g.~\cite{cosmoweinberg}and references therein). The 
decoherence of inflaton oscillations and other fields which are produced by the decay of the 
inflation oscillations make particles to behave classically. At the same time this process 
decoheres the metric fluctuations because metric is treated as a secondary (dependent) field. 
The result is a classical distribution of free particles but the particles/fields which stay 
correlated. Therefore we can interpret $f^i$ as a classical distribution.

In the classical limit the evolution of $f^i (x, p)$ is governed by the Boltzmann 
equation~\cite{houriuhecr}:
\bea
p^{\mu}{\partial}_{\mu} f^{(i)}(x,p) - ({\Gamma}^{\mu}_{\nu\rho} p^{\nu} 
p^{\rho}) \frac {\partial f^{(i)}}{\partial p^{\mu}} & = & 
-({\mathcal A}(x,p) + {\mathcal B}(x,p,\varphi)) f^{(i)}(x,p) + 
{\mathcal C}(x,p,\varphi) + \nonumber \\
 & & {\mathcal D}(x,p,\varphi) + {\mathcal E}(x,p). \label {bolt}
\eea
\bea
{\mathcal A}(x,p) & = & {\Gamma}_i m_i.\label {idec}\\
{\mathcal B}(x,p) &= & \sum_j \frac {1}{(2\pi)^3 g_i} \int d\bar p_j 
f^{(j)}(x,p_j) A (s){\sigma}_{ij}(s).\label {absint}\\
{\mathcal C}(x,p) & = & \sum_j {\Gamma}_j m_j \frac {1}{(2\pi)^3 g_i} \int 
d\bar p_j f^{(j)}(x,p_j)\frac {d{{\mathcal M}^{(i)}}_j}{d\bar p}.
\label {jdec}\\
{\mathcal D}(x,p) & = & \sum_{j,p}\frac {1}{(2\pi)^6 g_i}\int d\bar p_j 
d\bar p_k f^{(j)}(x,p_j)f^{(p)}(x,p_k) A (s) \frac {d{\sigma}_{j+p \rightarrow 
i+\ldots}}{d\bar p}. \label {proint} \\
A (p_i,p_j) & = & ((p_i.p_j)^2 - m_i^2m_j^2)^{\frac {1}{2}} = 
\frac {1}{2} ((s - m_i^2 - m_j^2)^2 - 4 m_i^2m_j^2)^{\frac {1}{2}}.
\label {kin}
\eea
where $m_i$, $\Gamma_i= 1/\tau_i$, $\tau_i$ are respectively mass, decay width, and lifetime of 
species $i$; ${\sigma}_{ij}$ is the total interaction cross-section between species $i$ and $j$ 
at a fixed center of mass energy $s$; $d{\sigma}_{j+k \rightarrow i+\ldots}/d\bar p = (2\pi)^3 E 
d\sigma/g_i p^2 dp d\Omega$ is the Lorentz invariant differential cross-section of production 
of $i$ in the interaction of $j$ and $k$; $g_i$ is the number of internal degrees of 
freedom (e.g. spin, color, etc.); $d\bar p = d^3p/E$ is the Lorentz invariant measure in 
momentum space; the term $d{{\mathcal M}^{(i)}}_{j}/d\bar p$ is the differential multiplicity 
of species $i$ in the decay of $j$; and finally ${\Gamma}^{\mu}_{\nu\rho}$ is the Levi-Civita 
connection. 
Note that the right hand side of (\ref{bolt}) is written in the local Minkovski frame in which 
the expression for the cross-sections is simple. As the effective mass of $\phi$ depends on 
the classical field $\varphi$, the right hand side of (\ref{bolt}) and thereby distributions 
$f^i (x, p)$ depend on $\varphi$. Thus, as mentioned at the beginning of this section, 
evolution of quantum and classical components of this model are coupled. 

Finally we must add Einstein equations to the set of evolution equations discussed above. 
We only consider scalar fluctuations in the linear regime and assume that the deviation of 
$f^{(i)}$ from a perfect fluid is small. With these simplifications the metric in Newtonian 
gauge can be written as:
\be
ds^2 = (1+2\psi (x)) dt^2 - a^2(t)(1-2\psi (x){\delta}_{ij}dx^idx^j = 
a^2(\eta) [(1+2\psi (x)) d{\eta}^2 - (1-2\psi (x) {\delta}_{ij}dx^idx^j)], 
\quad dt \equiv a d{\eta} \label{metric}
\ee
where $t$ and $\eta$ are respectively comoving and conformal times. Einstein equations 
for this metric can be found in textbooks, see for instance~\cite{cosmoweinberg}. We do not 
reproduce them here because in the present work we do not solve them along with 
field equations. Nonetheless we only remind the evolution equation for $a(t)$ which its 
evolution has special importance for dark energy:
\be
H^2 \equiv \frac{\dot{a}^2}{a^2} = \frac{8\pi G}{3}\sum_i \rho_i (t) \label{hubbleeq}
\ee
where $\rho_i$ is the energy density of species $i$. During radiation domination epoch the 
density of non-relativistic particles such as $X$ is by definition negligible, and 
evolution of $a (t)$ is governed by relativistic species which are not considered here 
explicitly. From the observed density of dark energy we can conclude that in this 
epoch its density was much smaller than other components, and had negligible effect on the 
evolution of expansion factor. In the matter domination epoch both $X$ and $A$ are assumed 
to be non-relativistic. If the lifetime of $X$ is much shorter than the age of the Universe 
at the beginning of matter domination epoch, most of $X$ particles have decayed, and 
it does not play a significant role in the evolution of $a (t)$ which is determined by other 
non-relativistic species. If the lifetime of $X$ is much larger than the age of the Universe, 
then $X$ particles can have a significant contribution in the total density of matter. As it 
decays very slowly, in the calculation of $a (t)$ it can be approximately treated as stable. 
In this case the evolution of $a (t)$ would be similar to a CDM model. A better estimation of 
$a (t)$ can be obtained by taking into account the decay of $X$ to relativistic particles. 
This method is used Ref.~\cite{houriustate} and $a (t)$ is calculated. However, here we use 
the simpler approximation because the problem in hand is very complex and we want to keep 
$a (t)$ decoupled from other equations.
At late times when the density of condensate becomes comparable to matter density the full 
theory, including Boltzmann equations must be solved. In this case the evolution of $a (t)$ 
is not simple and needs a full numerical solution and we leave it to a future work.

Quantum interactions happen at high energy scales, i.e. short distances. Thus, in the 
dynamic equations of the quantum fields we use only the homogeneous metric. This 
approximation is valid if at the epoch just after the production of $X$ particles $H \ll m_X$. 
Assuming a low energy preheating temperature $\sim 10^7$ GeV and $m_X \sim \Lambda_{GUT} 
\lesssim 10^{16}$ GeV, $m_X$ would fulfill the above condition. However, at early times this 
condition is not satisfied by the self-interacting field $\Phi$ because its mass is 
expected to be very small. In Appendix (\ref{app:d}) we show that at linear order the 
propagator of the fields in a background with small fluctuations are simply
$G_h(x,y)(1+\psi)$. Therefore we can use the homogeneous background to solve dynamic 
equations and then correct it for the effect of small fluctuations if necessary. For this 
reason in the following we mainly use a homogeneous background and discuss the effect 
of fluctuations afterward. Apriori the evolution equation of the condensate also must be 
written for a fluctuating metric, specially if $\Phi$ presents the quintessence field of a dark 
energy model. In Appendix (\ref{app:d}) this equation is determined in Newton gauge. However, 
due to the non-linearity of this equation, it is not possible to find a simple perturbative 
correction of the solution. Therefore in the present work we neglect metric fluctuations in 
this equation.

\section {Solution of the evolution equations} \label{sec:solevol}
In the previous section after writing all the dynamic equations it became clear that they 
are coupled and in general non-linear. In this situation it is impossible to proceed 
analytically and obtain a solution for these equations unless we break the mutual coupling 
of equations by taking some simplifying approximations. This should be possible because 
both mass and couplings of $\Phi$ are considered to be small. Therefore, the contribution 
of the condensate $\varphi$ in the evolution of particle distributions, equation (\ref{bolt}), 
must be small. Moreover, in this work our focus is on the formation of the condensate and an 
approximate distribution for other particles should be enough for a zero order estimation 
of the condensate evolution. Therefore, in place of solving the complete set of Boltzmann 
equations, we consider initial thermal distributions for the particles and assume that the 
interaction terms in the right hand side of (\ref{bolt}) leads to a slight difference 
between the effective temperature of species. This simplification is similar to temperature 
shift considered for the CMB in the treatment of SZ effect. Giving the fact that uncertainties  
on the initial temperature of the species is large, a slight modification should not have 
large effect on the condensation process and the properties of the condensate, otherwise the 
model would need fine-tuning and lose its reliability.

The advanced and retarded propagators defined in (\ref{proggt}) and (\ref{progls}), are 
needed for the calculation of expectation values in (\ref{valxaa}-\ref{valxaphi}). They 
are calculated on the non-vacuum states of the matter content of the Universe. Appendix 
\ref{app:a} explains in details the derivation of such propagators from vacuum Green's 
functions. Energy-momentum distributions of particles discussed above are necessary at this 
step. They make the connection between macroscopic cosmological evolution of the distribution 
of species and micro-physics of the condensate.

To solve the Green's functions and the evolution equation of the condensate, it is more 
convenient to write them with respect to conformal time $\eta$ along with a redefinition of 
$\varphi$:
\be
\chi \equiv a\varphi \label{chidef}
\ee
Then, if we considering only the homogeneous component of metric (\ref{metric}), the 
evolution equation of the classical field $\chi$ for the three decay models in 
(\ref{decaymode}) are the followings:
\bea
{\chi}'' - \delta_{ij}\partial_i\partial_j \chi + (a^2 m_{\Phi}^2 - 
\frac{a''}{a}) \chi + \frac{\lambda}{n}\sum_{i=0}^{n-1} a^{3-i}(i+1) 
\binom{n}{i+1} {\chi}^i \langle {\phi}^{n-i-1}\rangle - a^3 \mg \langle 
XA \rangle_a = 0 && \nonumber\\
\hspace{12cm} \text{For (\ref{decaymode})-a} && \label{evola} \\ 
{\chi}'' - \delta_{ij}\partial_i\partial_j \chi + (a^2 m_{\Phi}^2 - 
\frac{a''}{a})\chi + \frac{\lambda}{n}\sum_{i=0}^{n-1} a^{3-i}(i+1) 
\binom{n}{i+1} {\chi}^i \langle {\phi}^{n-i-1}\rangle - a^3 \mg \langle 
XA^2 \rangle = 0 && \nonumber\\
\hspace{12cm} \text{For (\ref{decaymode})-b} && \label{evolb} \\ 
{\chi}'' - \delta_{ij}\partial_i\partial_j \chi + (a^2 m_{\Phi}^2 - 
\frac{a''}{a})\chi + \frac{\lambda}{n}\sum_{i=0}^{n-1} a^{3-i} (i+1) 
\binom{n}{i+1} {\chi}^i\langle {\phi}^{n-i-1}\rangle - 2\mg a^2 \chi \langle 
XA \rangle_c - 2\mg a^3 \langle \phi XA \rangle = 0 && \nonumber\\
\hspace{12cm} \text{For (\ref{decaymode})-c} && \label {evolc}
\eea
Note that even if we neglect the expectation values related to the interaction 
between $X$, $A$ and $\phi$, due to the interaction of classical component (the condensate) 
with the quantum component the effective potential of the condensate $\varphi$ is not equal to 
the self-interaction term in the original Lagrangian and includes quantum corrections. 
In Sec. \ref{sec:classevol} we discuss the circumstances in which these corrections play 
an important role in the evolution of $\varphi$.

The vacuum propagators of quantum fields $\phi$, $X$ and $A$ are\footnote{Moat of the calculations in this work are performed on a space-like 3-surface. However, to simplify the notation, we omit the vector sign on 3-vectors except in places where this can make a confusion.}:
\bea
&& \frac{d^2}{d{\eta}^2} G^{\Upsilon}_F (x,y) - \delta_{ij}\partial_i\partial_j 
G^{\Upsilon}_F (x,y) + (a^2 m_{\Phi}^2 - \frac{a''}{a} + (n-1) \lambda 
a^{4-n}{\chi}^{n-2}) G^{\Upsilon}_F (x,y) = -i \frac {{\delta}^4 (x-y)}{a} 
\label {propup} \\
&&\frac{d^2}{d{\eta}^2} G^{\mathtt i}_F (x,y) - \delta_{ij}\partial_i\partial_j 
G^{\mathtt i}_F (x,y) + (a^2 m_{\mathtt i}^2 - \frac{a''}{a}) G^{\mathtt i}_F 
(x,y) = -i \frac {{\delta}^4 (x-y)}{a} \quad {\mathtt i} = \cx, \ca 
\label {propcx}\\
&& G^{\Upsilon}_F \equiv a(\eta)G^{\phi}_F \quad , \quad G^{\cx}_F \equiv a(\eta) G^X_F 
\quad , \quad G^{\ca}_F \equiv a(\eta) G^A_F \label {progredef}
\eea
with $f' \equiv df/d\eta$. Note that $\ca$, $\cx$ and $\Upsilon$ indices are respectively 
another name for $A$, $X$ and $\phi$ used only in the modified propagators which are defined 
in (\ref{progredef}).
The classical field $\chi$ (or equivalently $\varphi$) contributes in the mass term of the 
quantum component $\Upsilon$ (or equivalently $\phi$). This term couples the Green's function 
equation of $\phi$ to the evolution equation of condensate i.e. to one of the equations 
(\ref {evola}-\ref {evolc}) depending on the decay model. As the coupling $\lambda$ is assumed 
to be small, we can linearize (\ref{propup}) and use a WKB-like prescription to obtain an 
approximate solution. However, for solving evolution equations (\ref{evola}-\ref{evolc}) the 
linearization is not always a good approximation. In Sec. \ref{sec:classevol} we will argue 
that a late time non-zero slowly varying condensate can be obtained only when the full 
non-linear equations are considered. 

\subsection {Homogeneous solution of field equations} \label{sec:homosol}
The $X$ particles are presumably produced during reheating epoch~\cite{sdmprod} and their 
decay begins afterward. In this epoch relativistic particles dominate the energy density of 
the Universe, thus we first consider this epoch. Fortunately, in this epoch the homogeneous 
field equations have exact solutions. In matter domination epoch only for special cases an 
analytical solution exists. They are discussed in the Appendix \ref{app:c}. 

The expansion factor $a (\eta)$ in the radiation dominated epoch has the 
following time dependence:
\be
a = a_0 \biggl (\frac{t}{t_0} \biggr )^{\frac{1}{2}} = a_0 \frac{\eta}{\eta_0} \quad 
\Longrightarrow \quad a'' = 0
\label {expan}
\ee
After taking the Fourier transform of the spatial coordinates and neglecting the 
$\varphi$-dependent term in (\ref {propup}), the solutions of the associated homogeneous 
equation of (\ref{propup}) and 
(\ref{propcx}) are well known~\cite{qmcurve}~\cite{integbook}:
\bea
&& (\frac{d^2}{d{\eta}^2} + k^2 + a^2 m_i^2) \um_k^i (\eta) = 0 
\label{eqhomo} \\
&& \um_k^i = \int d^3 \vec{x} \um^i (x) e^{i\vec{k}.\vec{x}} \quad i = 
\Phi~,~\cx~,~\ca \label {fourier}\\
&& \um_k^i (\eta) = c_k^i U_k (i\alpha_i, z' e^{i\frac{\pi}{4}}) + 
d_k^i V_k (i\alpha_i, z' e^{i\frac{\pi}{4}}) \label{homosol} \\
&& z' \equiv \theta_i \frac{\eta}{\eta_0}, \quad \theta_i \equiv \sqrt{2 a_0 \eta_0 m_i} = 
\sqrt{\frac{2m_i}{H_0}} \quad \alpha_i \equiv \frac{k^2\eta_0}{2 a_0 m_i} = 
\frac{k^2 H_0\eta_0^2}{2m_i} \label{homosolparam}
\eea
where $a_0$ and $H_0$ are respectively the expansion factor and Hubble constant at 
initial conformal time $\eta_0$. The functions $U_k$ and $V_k$ are two independent 
parabolic cylindrical functions~\cite{parabcylinder}:
\bea
U (a,z) &=& y_1 \cos \pi (\frac{1}{4} + \frac{a}{2}) - y_2 \sin \pi (\frac{1}{4} + 
\frac{a}{2}) \label{parbcylu} \\
V (a,z) &=& y_1 \sin \pi (\frac{1}{4} + \frac{a}{2}) + y_2 \cos \pi (\frac{1}{4} + 
\frac{a}{2}) \label{parbcylv} \\
y_1 (a,z) &=& \frac{\Gamma (\frac{1}{4} - \frac{a}{2})}{\sqrt{\pi} 2^{\frac{1}{4} + 
\frac{a}{2}}} e^{-\frac{z^2}{4}}~_1F_1 (\frac{1}{4} + \frac{a}{2};\frac{1}{2};
\frac{z^2}{2}) \label{parbcylyone} \\
y_2 (a,z) &=& \frac{\Gamma (\frac{3}{4} - \frac{a}{2})}{\sqrt{\pi} 2^{\frac{1}{4} + 
\frac{a}{2}}} e^{-\frac{z^2}{4}}~_1F_1 (\frac{3}{4} + \frac{a}{2};\frac{3}{2};
\frac{z^2}{2}) \label{parbcylytwo}
\eea
From now on for simplicity we drop the species index $i$ except when its presence is 
necessary. We call two independent solutions of (\ref{eqhomo}) in a general basis $\um_k$ 
and $\vm_k$. If we want that these solutions correspond to the coefficients of the canonical 
decomposition of $\phi$, equation (\ref{canon}) in Appendix \ref{app:a}, we must 
choose a basis such that $\vm_k = \um_k^*$. In the rest of this work we only consider this 
basis. Note that the two solutions $U$ and $V$ in (\ref{homosol}) are not complex conjugate 
of each other and therefore can not be identified with  $\um_k$ and $\vm_k$.

We are interested on the asymptotic behaviour of $U$ and $V$ functions when $\eta/\eta_0 
\gg 1$. Their asymptotic expressions are (see e.g. ~\cite{paracylasymp} and 
references therein):
\bea
U_k (i\alpha_i, z' e^{i\frac{\pi}{4}}) &=& e^{\frac{\pi\alpha_i}{4} - 
i\frac{\pi}{8}} {z'}^{-\frac{1}{2}} e^{-i(\alpha_i \ln z' 
+ \frac{{z'}^2}{4})} \sum_{s=0}^\infty (2i)^s (\frac{1}{4} + 
\frac{i\alpha_i}{2})_s~\frac{(\frac{i\alpha_i}{2} + \frac{3}{4})_s}{s! {z'}^2s} 
\label{uasymp} \\
V_k (i\alpha_i, z' e^{i\frac{\pi}{4}}) &=& iU_k (i\alpha_i, z' e^{i\frac{\pi}{4}}) + 
\sqrt {\frac{2}{\pi}} \Gamma (-i\alpha_i + \frac{1}{2}) 
e^{-\frac{\pi\alpha_i}{4} - i\frac{\pi}{8}} {z'}^{-\frac{1}{2}} 
e^{i(\alpha_i \ln z' + \frac{{z'}^2}{4})} \nonumber \\
&& \sum_{s=0}^\infty (-2i)^s (\frac{\frac{1}{4} - i\alpha_i}{2})_s~\frac{(\frac{3}{4} - 
\frac{i\alpha_i}{2})_s}{s! {z'}^{2s}} \\\label{vasymp}
(b)_s &\equiv &\frac{(b+s-1)!}{(b-1)!}
\eea

The evolution equation for free Feynman propagators - the 2-point Green's functions - 
is defined as:
\be
(\frac{d^2}{d{\eta}^2} + k^2 + a^2 m_i^2) G_k (\eta,\eta') = 
-i \frac {\delta (\eta - \eta')}{a} \label{prophomo}
\ee
When $\eta \neq \eta'$, equation (\ref{prophomo}) is the same as the 
homogeneous equation (\ref{eqhomo}), and therefore the solution of 
(\ref{prophomo}) is a linear combination of two independent solutions of 
(\ref{eqhomo}). According to the definition of Feynman propagators 
(\ref{proggt}) and (\ref{progls}), they can be divided to advanced and retarded 
propagating components $G^<$ and $G^>$. The transformation $\eta \leftrightarrow \eta'$ 
changes the role of these propagators, $G^< \leftrightarrow G^>$. After adding the effect of a 
non-vacuum state, as explained in Appendix \ref{app:a}, the non-vacuum 
propagator $G (\eta, \eta')$ has the following expansion:
\bea
iG_k (\eta, \eta') &=& \biggl [{\mathcal A}^>_k \um_k (\eta)\um^*_k 
(\eta') + {\mathcal B}^>_k \um^*_k (\eta) \um_k (\eta')\biggr ] \Theta 
(\eta - \eta') + \nonumber \\
&& \biggl [{\mathcal A}^<_k \um_k (\eta)\um^*_k (\eta') + 
{\mathcal B}^<_k \um^*_k (\eta) \um_k (\eta')\biggr ] \Theta (\eta' - \eta) 
\label{progexpand}
\eea
where ${\mathcal A}^>_k$, ${\mathcal B}^>_k$, ${\mathcal A}^<_k$ and 
${\mathcal B}^<_k$ are integration constants. In the Appendix \ref{app:a} 
we show that for the free propagators - at the lowest perturbation order - 
if the state $|\Psi\rangle$ is not vacuum, it is possible to 
include its effect in the boundary conditions imposed on the propagator. 
Comparing (\ref{progexpand}) with (\ref{propst}) in Appendix \ref{app:a}, 
the relation between these constants and the initial state can be found:
\bea
&&{\mathcal A}^>_k = 1 + {\mathcal B}^>_k \quad , \quad {\mathcal B}^<_k = 
1 + {\mathcal A}^<_k \label{abrel} \\
&&{\mathcal A}^<_k = {\mathcal B}^>_k = \sum_i \sum_{k_1 k_2 
\ldots k_n} \delta_{kk_i} |\Psi_{k_1 k_2 \ldots k_n}|^2 \label{abpsi}
\eea
See Appendix \ref{app:a} for a detailed description of $|\Psi|^2$. It is easy 
to see that with relations (\ref{abrel}) and (\ref{abpsi}) between 
constant coefficient, the consistency condition defined as:
\be
G_k^> (\eta,\eta')\biggl |_{\eta = \eta'} = G_k^< (\eta,\eta')\biggl |_{\eta = 
\eta'} \label {consistcond} 
\ee
is automatically satisfied. Therefore propagators over a non-vacuum state 
$\Psi$ depend only on this state and the solutions of the field equation. 
They depend also on two arbitrary constants $c_k^i$ and $d_k^i$ in the solution of 
field equation. They must be fixed by the initial conditions too.

There is one more consistency condition that propagators must satisfy. 
By integrating the two sides of the equation (\ref{prophomo}) with respect 
to $\eta$ in an infinitesimal region around $\eta'$ and by using the solution 
(\ref{progexpand}), we find the following constraint:
\be
\um_k^{'} (\eta) \um^*_k (\eta) - \um_k (\eta) \um^{'*}_k (\eta)= 
\frac {-i}{a(\eta)} \label{derivcond}
\ee
This relation fixes one of the integration constants in (\ref{homosol}). It is easy 
to see that a constant shift of the argument of $\um_k (\eta)$ does not violate 
(\ref {derivcond}). This means that the phase of $\um_k (\eta)$ is not an observable 
and can be fixed arbitrarily. It is also interesting to note that multiplication of 
two sides of (\ref{derivcond}) with an arbitrary constant rescales $a(\eta)$ which is 
equivalent to redefinition of $a_0$. Rescaling of $a_0$ is equivalent to redefinition 
of coordinates and therefore is not an observable. In Minkovski spacetime the scale 
factor $a$ is fixed to 1, and therefore there is no place for rescaling. In another 
word, in a Minkovski space the normalization of the propagators is an observable and 
affects the final results. This scaling property in FLRW and De-Sitter metrics is a 
consequence of diffeomorphism invariance in the framework of curved spacetimes and 
general relativity.

\subsection{Initial conditions for propagators} \label{sec:initcond}
Field equations are second order differential equations and need the initial value of the 
field and its derivative or a combination of them to obtain a complete description of the 
solutions. The general initial conditions for a bounded system - including both Neumann and 
Dirichlet conditions as special cases - are the followings~\cite{qftinit}:
\be
n^\mu \partial_\mu \um = -i{\mathcal K} \um \quad ,\quad g_{\mu\nu}n^\mu n^\nu = 1 
\label{geninit} 
\ee
The 4-vector $n^\mu$ is the normal to the boundary surface. If the boundary is space-like, 
$n^\mu$ can be normalized as $n^{\mu} = a^{-1}(1,0,0,0)$. Then:
\bea
a^{-1}\partial_{\eta} \um = -i {\mathcal K} \um \label{geninitorth} 
\eea
The constant ${\mathcal K}$ depends on the scale $k$. In any boundary problem, the 
boundary conditions must be defined for all the boundaries. Thus, in a cosmological setup 
the initial condition constraints (\ref{geninitorth}) must be applied to both past (initial) 
and future (final) boundary surfaces~\cite{qftinit}. But in the case of propagators, they are 
respectively applicable to past and future propagators only. In each case the other boundary 
condition is the consistency condition (\ref{consistcond}). Assuming different values for 
${\mathcal K}$ on these boundaries, we find:
\be
{\mathcal K}_j = i\frac {\um^{'}_k (\eta_j)}{a_j \um_k(\eta_j)} \quad , 
\quad j = i, f 
\label {alphavac}
\ee
Indexes $i$ and $f$ refer to the value of quantities at initial and final 3-surfaces. These 
boundary conditions relate ${\mathcal K}_i$ and ${\mathcal K}_f$ to $c_k$ and $d_k$ in 
(\ref{homosol}). In fact using (\ref{derivcond}) along with (\ref{homosol}) we find:
\be
|\um_k (\eta_j)|^2 = \frac {1}{a^2(\eta_j) ({\mathcal K}_j (k,\eta_j) + 
{\mathcal K}^*_j (k,\eta_j))} \quad , \quad |\um^{'}_k (\eta_j)|^2 = 
\frac {|{\mathcal K}_j (k,\eta_j)|^2}{{\mathcal K}_j (k,\eta_j) + 
{\mathcal K}^*_j (k,\eta_j)} \quad j = i, f
\label{uband}
\ee 
Application of (\ref{uband}) to the solution (\ref{homosol}) at two boundaries fixes 
dynamical constant $c_k$ and $d_k$ as functions of ${\mathcal K}_i$ and ${\mathcal K}_f$ up 
to a constant phase. In fact as we mentioned previously, due to the equality (\ref{derivcond}) 
the phase of $\um_k$ is not an observable. Therefore, we can assume that it is zero on the 
initial and final boundaries. In the next section we will calculate propagators and evolution 
equation of $\chi$ in radiation and matter domination epoch separately. Thus, the initial and 
final boundaries correspond to the beginning and end of each epoch, and we must respect 
continuity condition, i.e. the initial condition for one epoch corresponds to the end 
condition for the previous epoch. 

In a cosmological context ${\mathcal K}_f$ can be decided based on observations, but 
${\mathcal K}_i$ is unknown and leaves one model dependent constant that should be fixed 
by the physics of early universe and the special state of our Universe among all possible 
states. This arbitrariness of the general solution or in other words the vacuum of 
the theory is well known~\cite{vacuu}. In the case of inflation, a class of 
possible vacuum solutions called $\alpha$-vacuum are usually used:
\be 
{\mathcal K}_i , {\mathcal K}_f = \sqrt {k^2/a^2_{i,f} + m^2} 
\label{bunchdavies}
\ee
and one obtains the well known Bunch-Davies solutions~\cite{qftinit}. 

Alternatively one can fix the solution $\um_k$ at one of the boundaries and apply the 
boundary condition (\ref{alphavac}) only to that 3-surface. Although this does not solve the 
problem of arbitrariness of ${\mathcal K}$ and its $k$ dependence, it reduces it to only one 
of the boundary surfaces, for instance to the final 3-surface. This make the choice of 
(\ref{bunchdavies}) physically motivated. In addition, by applying (\ref{alphavac}) only to 
one of the boundaries, the causality of the constraint (\ref{geninitorth}) is more 
transparent and the state of the second boundary is directly related to the physics of the 
first one through the evolution equation. We remind that the evolution of the expansion factor 
$a (\eta)$ is related to all type of matter including the condensate through the equation 
(\ref{hubbleeq}).

An additional and somehow hidden arbitrariness in this formalism is the fact that apriori 
$k$ dependence of the boundary constant ${\mathcal K}$ does not need to be the same for all 
the fields of the model. However, different $k$ dependence breaks the Equivalence Principal. 
Similarly, a value different from (\ref{bunchdavies}) for ${\mathcal K}$ will lead to the 
breaking of the translation symmetry~\cite{qftinit}~\cite{infrenorm}. In the context of 
quantum gravity the violation of both these laws are expected and therefore, in a general 
framework the choice of different ${\mathcal K}$ for fields is allowed.

Before finishing this section we discuss also $k$-dependence of $U_k$ and $V_k$. This is 
specially important in the context of dark energy because no strong fluctuation is 
observed in its spatial distribution. The solutions of equation (\ref{eqhomo}) depends on 
$k$ only through $\alpha_i$ defined in (\ref{homosolparam}). From the asymptotic expression 
of $U_k$ it is evident that apart from a constant term which can be absorbed in the 
integration constants, other terms containing $\alpha_i$ are either oscillating or 
approach a constant for $\alpha_i \gg 1$ which is equivalent to $|k| \rightarrow \infty$.
It can be shown that $V_k$ also has the same type of behaviour because 
$|\Gamma (-i\alpha_i + 1/2 |^2 = \pi/\cosh \pi \alpha_i$. As for the integration constants 
$c_k$ and $d_k$, from (\ref{uband}) and (\ref{bunchdavies}) it is easy to see that for large 
$|k|$ they are proportional to $1/\sqrt{|k|}$. Thus the amplitude of the Green's functions 
is asymptotically proportional to $1/|k|$ and an oscillatory $k$ dependent component.

\subsection{Propagator of light scalar} \label{sec:wkbprog}
Finally after finding the solution of free propagators (\ref{propup}-\ref{progredef}) 
without $\varphi$ dependent terms, we use a WKB-like approximation to correct the 
propagator of $\phi$ for the contribution of self-interaction in its mass. We neglect 
the effect of varying mass on $\alpha_\phi$ and replace $z'$ in $\um$ with:
\be
z' \rightarrow \sqrt{2a_0\eta_0 m} \int d(\frac{\eta}{\eta_0}) \biggl (1 + (n-1) 
\lambda \frac{\varphi_k^{n-2}(\eta)}{m_{\Phi}^2} \biggr )^{\frac{1}{4}} 
\label{wkbappr}
\ee
where $\varphi_k$ is the Fourier transform of $\varphi (x)$ (see Appendix \ref{app:a} for 
the details of approximations) and $\alpha_\Phi$ is defined in (\ref{homosolparam}).

With this correction we have the solution of all the free Green's functions at lowest 
order. However, the dependence of $G_k^{\phi}$ on $\varphi (x)$ (or equivalently $\chi$) 
couples propagators to the evolution equation of $\varphi$ i.e one of the equations 
(\ref{evola}), (\ref{evolb}) or (\ref{evolc}), depending on the interaction mode. A complete 
solution can be obtained only through numerical calculations. Nonetheless, when we calculate 
expectation values, the correction (\ref{wkbappr}) produces high order of $\varphi$ in a 
polynomial expansion which at lowest order of approximation can be neglected. This simplifies 
analytical tracking of the evolution of condensate.

\section{Evolution of the condensate} \label{sec:classevol}
Finally in this section we study the evolution of the condensate during the cosmic time from 
the end of massive production of $X$ particles - presumably the end of reheating - until matter 
domination epoch. Our aim is to find analytical approximations for the solutions of equations 
(\ref{evola}-\ref{evolc}) which rule the evolution of the condensate for the three 
decay/interaction models considered in this work. We describe the solution for the model (a) 
which is the simplest one in details. For other two models we briefly explain their differences 
and deviations from model (a). 

We begin with the radiation domination epoch and use the results obtained in Sec. 
\ref {sec:solevol}. Then, we consider the matter domination epoch for which the solutions of 
field equation are discussed in Appendix \ref{app:c}. Before going to details we should make a 
remark about the necessity of considering quantum corrections in the evolution of the 
condensate. The reason is the complexity of equations (\ref{evola}-\ref{evolc}) and thereby 
the solutions obtained in this section. They raise a question about the necessity of 
considering such complex model and the significance of quantum corrections in the evolution 
of condensate. After all, in other contexts such as inflation, a classical homogeneous field 
is considered and only fluctuations around it are quantized and studied. We should remind 
that here we are considering a situation in which initially the scalar field is uniformly 
null. Such an initial point is the minimum of a $\Phi^n$, $n > 0$ potential which has a clear
physical interpretation in many-body quantum systems and perturbative field theories. 
Spinorial models in which the potential has a non-zero maximum at origin induce an imaginary 
(tachyonic) mass term and can not correspond to a fundamental field and must be considered 
as an effective field. The same argument applies to inverse power-law potentials which are 
present in the majority of quintessence models. In fact, in all these cases the potential 
is the effective potential of the scalar condensate not the fundamental quantum 
field itself - the best example of such models is Higgs. Therefore, when we solve the dynamic 
equation with this minimum point as initial condition, no condensate can form unless an 
additional (quantum) interaction is present. In the case of inflation, this 
pre-inflationary step is usually overlooked and its presence is simply assumed as an initial 
condition. Therefore, the issue of the formation of a cosmic scalar - a condensate - discussed 
here is relevant for inflation models too.

\subsection {Radiation domination epoch} \label{sec:raddom}
Model (a) is the simplest case between models considered in this work. We first neglect 
the self-interaction term in (\ref{evola}) and take the Fourier transform of the left hand 
side of this equation (See Appendix \ref{app:a} for technical details and approximations that 
have been taken):
\bea
&& {\chi}'' + (k^2 + a^2 m_{\Phi}^2) \chi + \frac{i \mg^2 a \chi (k)}{(2\pi)^6} 
\int d^3 k_1 d^3 k_2 \delta^{(3)} (\vec{k} - \vec{k}_1 - \vec{k}_2) \nonumber \\
&& \quad\quad \int d\eta'\sqrt{-g} \biggl [G_{k_1}^{\ca >} (\eta,\eta') G_{k_2}^{\cx >} 
(\eta,\eta') - G_{k_1}^{\ca <} (\eta,\eta') G_{k_2}^{\cx <} (\eta,\eta') \biggr ] = 0 
\label {aint}
\eea
This equation is linear and we can use WKB-like methods to find an approximative solution.
If we neglect interaction terms, equation (\ref{aint}) has an exact solution of the form 
(\ref{homosol}). Because it is assumed that the coupling $\mg$ is small\footnote{In model 
(a) the coupling $\mg$ has a mass dimension of 1. Therefore the statement about the 
weakness of interaction in (\ref{aint}) or in other word the smallness of $\mg$ refers to its 
comparison with $(k^2+a^2 m_\Phi^2)$ term.}, a WKB-like methods allows to find an 
approximate analytical solution when the interaction terms are taken into account:
\bea
\chi_k^{(a)} (\eta) &=& c_k^{(a)} U_k (i\alpha_\Phi, \wm_k^{(a)} (\eta) \theta_\Phi 
e^{i\frac{\pi}{4}}) + d_k^{(a)} V_k (i\alpha_\Phi, \wm_k^{(a)} (\eta) \theta_\Phi 
e^{i\frac{\pi}{4}}) \label{amodelsolr} \\
\wm_k^{(a)} (\eta) & \equiv & \int d\eta \biggl \{1+ \frac{i \mg^2}{(2\pi)^3 m_\Phi^2 
a(\eta)} \int d^3 k_1 d^3 k_2 \delta^{(3)} (\vec{k} - \vec{k}_1 - \vec{k}_2) \nonumber \\ 
&& \int d\eta'\sqrt{-g} \biggl [G_{k_1}^{\ca >} (\eta,\eta') G_{k_2}^{\cx >} 
(\eta,\eta') - G_{k_1}^{\ca <} (\eta,\eta') G_{k_2}^{\cx <} (\eta,\eta') \biggr ]\biggr \}
\label{amodelwkbr} 
\eea
where the index $(a)$ refers to the interaction model. Considering the complexity of 
expressions (\ref{propfuphi}) and (\ref{proppaphi}) for propagators and multiple integrals 
in (\ref{amodelwkbr}), it is evident that the expression (\ref{amodelsolr}) 
is very involved. But, we are essentially interested in the asymptotic behaviour of 
$\chi_k$ to see whether it grows with time. For this reason, in place of presenting the full 
expression of the integrals, we concentrate on growing terms, their order, and the conditions 
for their existence. 

Using equations (\ref{uasymp}) and (\ref{vasymp}), the asymptotic expression of the functions 
$U_k$ and $V_k$ respectively, we find that the asymptotic expression of $\chi_k$ includes 
terms of the following form:
\be
(z' + h_k (z'))^{-\frac{1}{2} \pm i\alpha_\Phi}~e^{\pm i (z' + h_k (z'))^2} \label {terms}
\ee
where $z',~\alpha_\Phi$, and $\theta_\Phi$ are defined in (\ref{homosolparam}) and 
$h_k (z') = \wm_k^{(a)} (\eta) \theta_\Phi - z' \ll z'$. We are interested in the large scale 
modes i.e. $k \ll m_\Phi$ for which the parameter $\alpha_\Phi$ is expected to be small. 
The power-law factor in (\ref{terms}) decays as $\sim 1/\sqrt{\eta}$. On the other hand, 
terms with in which $\pm i (z' + h_k (z'))^2$, the exponent of the exponential factor, has 
a positive real part will grow exponentially and ones with opposite sign will decline 
exponentially. Thus, in radiation domination epoch the production of $\Phi$ particles by 
the decay of $X$ alone is enough for the formation of a condensate. However, due to the 
smallness of the interaction term (\ref{amodelwkbr}) when $\tau \gg$ age of the Universe 
in that epoch, the growth of the condensate can be very slow. 

To complete this argument we must also show that the function $f_k (z')$, in particular 
its imaginary part is not zero. Therefore, we must determine the multiple integrals 
in (\ref {amodelwkbr}). Considering the complexity of propagators (\ref {propfuphi}) and 
(\ref {proppaphi}), it is evident that their calculation is long and tedious but 
straightforward. Thus, in place of presenting the complete expression of $f_k(z')$, 
we only explain the general form of the terms it contains and their behaviour for 
$\eta/\eta0 \gg 1$ which is relevant to the asymptotic evolution of condensate $\varphi$. 
We also use the asymptotic expressions of the functions $U_k$ and $V_k$ to expand and 
calculate the propagators in the closed time path integrals. In addition, as we discussed 
in Appendix \ref{app:a}, we consider that barred coordinates in the propagators 
(\ref {propfuphi}) and (\ref {proppaphi}) are independent from $x$ and $y$. They present the 
average space-time coordinates in which these processes occur. With these simplifications,  
thermal distributions in the propagators do not contribute in the integration over time 
in (\ref{amodelwkbr}). The integral over $\eta'$ includes terms of the following form:
\be
\int dx x^{3+i\alpha} ~ e^{i\frac{\beta x^2}{4}} = - (\frac{4i}{\beta})^{2+\frac{i\alpha}
{2}} ~ \Gamma (2+\frac{i\alpha}{2}, -\frac{i\beta}{4} x^2), \quad x \equiv 
\frac{\eta'}{\eta_0}  \label{etapint}
\ee
where parameters $\alpha = (\pm \alpha_A \pm \alpha_X)$ and $\beta = (\pm \theta_A^2 \pm 
\theta_X^2)$ (all combinations of signs are present). The integral in (\ref{etapint}) 
includes the contribution of both advanced and retarded propagators that cover distinct 
time domains - for a given $\eta$, in advanced term $\eta' < \eta$ and in retarded term 
$\eta' > \eta$. 

The next step is taking the indefinite integral over $\eta$ which includes terms of the 
following form:
\be
\int \frac {dx}{x} ~ \Gamma (2+\frac{i\alpha}{2}, -\frac{i\beta}{4} x^2) \approx 
-\frac{1}{2} ~ (\frac{\beta}{4})^{- \frac{i\alpha}{2}} ~ e^{\frac{\pi \alpha}{4}} ~ 
(\frac{\eta}{\eta_0})^{2\alpha i} ~ e^{- \frac{i\beta \eta^2}{4 \eta_0^2}}  \label{etaint}
\ee
For this integration we have used the asymptotic expression of incomplete Gamma functions 
$\Gamma (\alpha, x) \approx x^{\alpha - 1} e^{-x}$. Note that the $x^{-1}$ factor in the 
integrand is due to the $a^{-1}(\eta)$ factor in (\ref{amodelwkb}). It has an 
important role in the evolution of the condensate and its presence depends on the 
interaction model. Finally, we find the following expression for $\wm_k^{(a)} (\eta)$ when 
$\eta/\eta_0 \gg 1$: 
\be
\wm_k^{(a)} (\eta) \approx \eta + \frac{i \mg^2 \eta_0^2}{(2\pi)^3 m_\Phi^2} 
\int d^3 k_1 d^3 k_2 \delta^{(3)} (\vec{k} - \vec{k}_1 - \vec{k}_2) \sum_{\alpha,\beta} 
A_{\alpha\beta} (k_1,k_2,\bar{x})~\exp \biggl (i (2\alpha \ln \frac{\eta}{\eta_0} - 
\frac {\beta \eta^2}{4 \eta_0^2}) \biggr ) \label{wksol}
\ee
The factors $A_{\alpha\beta}$ depend on $\alpha_A$, $\alpha_X$, $\theta_A$, $\theta_X$, and 
thereby on $k_1$ and $k_2$. The parameter $\alpha$ also depends on the momentums, but it 
appears only in the logarithmic term which increases much slower than the quadratic 
term. Thus, its effect can be important only at large momentums. More importantly 
$A_{\alpha\beta}$ factors depend on the distributions of $X$ and $A$ particles. Each of them 
includes one of the following factors: $f^{(A)}f^{(X)}$, $f^{(A)}(1+ f^{(X)})$, 
$(1+f^{(A)})f^{(X)}$, or $(1+f^{(A)})(1+f^{(X)})$. This is the reason for the inclusion of 
$\bar{x}$ in (\ref{wksol}). As we mentioned before, we first consider $\bar{x}$ as an 
independent variable. Then, after the calculation of closed time path integrals, we identify 
it with 
$\eta$ (neglecting its spacial dependence). Note that due to the constant term in some of 
the above factors, the interaction term is not zero when distributions are null, i.e. when 
expectation values are calculated for vacuum. This apparent inconsistency can be solved if 
the initial growth rate of $\chi$ assumed to be null, which is consistent with the concept of 
vacuum as the absence of any particle, see Sec. \ref{sec:initcondcon} for details. Before 
proceeding with the final expression for $U_k$ and $V_k$, we discuss the distribution 
of $A$ and $X$ particles.

The $X$ particles are expected to be heavy with a mass of the order of GUT scale 
$\sim 10^{16}$~GeV. On the other hand, reheating temperature must be $\lesssim 10^9$~GeV~
to prevent overproduction of gravitinos~\cite{reheattemp}. Therefore, we expect that at 
production and thermalization epoch - if $X$ particles have ever been in thermal equilibrium 
with the rest of the Universe~\cite{uni,sdmprod} - their temperature or more generally their 
kinetic energy was much smaller than their mass. Thus, we can safely consider their 
temperature to be zero. This simplifies the expression of their propagators. In this case 
their density evolves as:
\be
\rho_X (\bar{x}) = \rho_X (\bar{x}_0) \frac{a_0^3}{a^3} e^{-\frac{t-t_0}{\tau}}, \quad\quad 
f^{(X)} (\bar{x},p) \approx \frac{2\pi^2 \rho_X(\bar{x})}{m_X} \delta (|p|)\label{xdens}
\ee
Using $\rho_X (\bar{x})$ and the effective temperature of $A$ particles, equation 
(\ref{atemprel}) in Appendix \ref{app:a}, we find that at $t \gg t_0$:
\be
{\mathsf k}_B T_A \approx (\frac{\pi^2 {\mathcal M}_A \rho_X (\bar{x}) t_0}{3 \zeta (4) 
\tau})^{\frac{1}{4}} e^{-\frac{t-t_0}{4\tau}}\label{atempapp}
\ee
This means that if $(t-t_0) \ll \tau$ during radiation domination, after a rapid rise of 
$T_A$ from zero, see (\ref{tempfrac}) the effective temperature of $A$ particles quickly 
approaches to a constant value. Therefore, their distribution during this epoch approaches 
$f^{(A)} \approx exp (-\beta_A \sqrt {k^2/a^2 + m_A^2})$ where $\beta_A$ is roughly constant. 
Due to the presence of $a^{-2}$ factor the effect of scale dependent term in $f^{(A)}$ 
decreases with time. Note that the effect of gravitational growth is included only in the 
effective temperature (\ref{atempapp}) because we didn't solve the Boltzmann equation for 
$A$. This should be an enough good approximation for the needs of the present work as 
we are mostly interested in large scale behaviour of the fields and distributions of 
particles.

A complete analytical expression of $\wm_k^{(a)}$ needs also the integration over momentums 
$k_1$ and $k_2$. One of these triple-integrals is canceled by the delta function presenting 
the conservation of momentum: $\vec{k_2} = \vec{k} - \vec{k_1}$. The other integral is 
reduced to a double integral because without loss of generality the $z$-axis can be selected 
to be orthogonal to the plane $\vec{k}-\vec{k_1}$. Nonetheless, these integrals are very 
complex except for the terms containing $f^{(X)}$. The presence of a delta functions 
in these terms, see equation(\ref{xdens}), reduces the second triple-integrals too. Fortunately 
these terms are much larger than terms due to the contribution of vacuum which are 
subdominant and can be neglected. However, equation (\ref{xdens}) shows also that $f^{(X)}$ 
decreases with the expansion of the Universe as well as with the decay of $X$ particles. 
In this case, at 
late times the vacuum terms can become dominant. Nonetheless, if a significant fraction 
of $X$ particles persist until the end of the radiation domination epoch, the terms 
proportional to $f^{(X)}$ continue to be dominant, and therefore $\wm_k$ has the following 
approximate expression:
\be
\wm_k^{(a)} (\eta) \approx \eta + \frac{i \mg^2 \eta_0^2}{(2\pi)^3 m_\Phi^2} 
\sum_{\alpha,\beta} A_{\alpha\beta} (k,\bar{x})~\exp \biggl (i (2\alpha \ln 
\frac{\eta}{\eta_0} - \frac {\beta \eta^2}{4 \eta_0^2}) \biggr ) \label{wksolint}
\ee
In Sec. \ref{sec:initcond} we showed that Green's functions are asymptotically proportional 
to $1/|k|$. Therefore, we expect that $\wm_k^{(a)} \propto 1/|k|^2$, and large $|k|$ 
modes decay quickly. This is consistent with the behaviour of dark energy.

After applying (\ref{wksolint}) to $U_k$ and $V_k$ in (\ref{amodelsolr}), we obtain - up to 
constants that we include in $c_k$ and $d_k$ - the following expressions:
\bea
U_k &\approx& \sqrt{\frac{\eta_0}{\eta}} \exp \biggl (\frac{1}{2} \sum_{\alpha,\beta}
B'_{\alpha\beta} \sin~(2\alpha \ln \frac{\eta}{\eta_0} - \frac {\beta \eta^2}
{4 \eta_0^2}) \biggl [\frac{\eta}{\eta_0} + A'_{\alpha\beta} \cos~(2\alpha \ln 
\frac{\eta}{\eta_0} - \frac {\beta \eta^2}{4 \eta_0^2}) \biggr ] \biggr ) \times \nonumber \\
&& \exp \biggl (-\frac{i}{4} \sum_{\alpha,\beta} \biggl \{ \biggl [\frac{\eta}{\eta_0} + 
A'_{\alpha\beta} \cos~(2\alpha \ln \frac{\eta}{\eta_0} - \frac {\beta \eta^2}{4 \eta_0^2}) 
\biggr ]^2 - {B'}^2_{\alpha\beta} \sin^2 (2\alpha \ln \frac{\eta}{\eta_0} - 
\frac {\beta \eta^2}{4 \eta_0^2}) \biggr \} \biggr ) \label{uapprox} \\
V_k - iU_k &\approx& \sqrt{\frac{\eta_0}{\eta}} \exp \biggl (-\frac{1}{2} 
\sum_{\alpha,\beta} B'_{\alpha\beta} \sin~(2\alpha \ln \frac{\eta}{\eta_0} - 
\frac {\beta \eta^2}{4 \eta_0^2}) \biggl [\frac{\eta}{\eta_0} + A'_{\alpha\beta} 
\cos~(2\alpha \ln \frac{\eta}{\eta_0} - \frac {\beta \eta^2}{4 \eta_0^2}) \biggr ] 
\biggr ) \times \nonumber \\
&& \exp \biggl (\frac{i}{4} \sum_{\alpha,\beta} \biggl \{ \biggl [\frac{\eta}{\eta_0} + 
A'_{\alpha\beta} \cos~(2\alpha \ln \frac{\eta}{\eta_0} - \frac {\beta \eta^2}{4 \eta_0^2}) 
\biggr ]^2 - {B'}^2_{\alpha\beta} \sin^2 (2\alpha \ln \frac{\eta}{\eta_0} - 
\frac {\beta \eta^2}{4 \eta_0^2}) \biggr \} \biggr ) \label{vapprox}
\eea
The presence of a real exponential term in both independent solutions of the evolution 
equation and the phase difference between them means that in the radiation domination 
epoch there is always a growing term that assures the accumulation of the condensate, 
although due to the smallness of the coefficients $A'_{\alpha\beta}$ which is proportional to 
$\mg^2$, its growth can be very slow. Therefore, we conclude that in this regime the 
production of $\Phi$ particles by the slow decay of $X$ particles according to 
the model (a) is enough to produce a growing condensate. 

\begin{figure}
\begin{center}
\begin {tabular}{ccc}
\includegraphics[height=5cm,angle=-90]{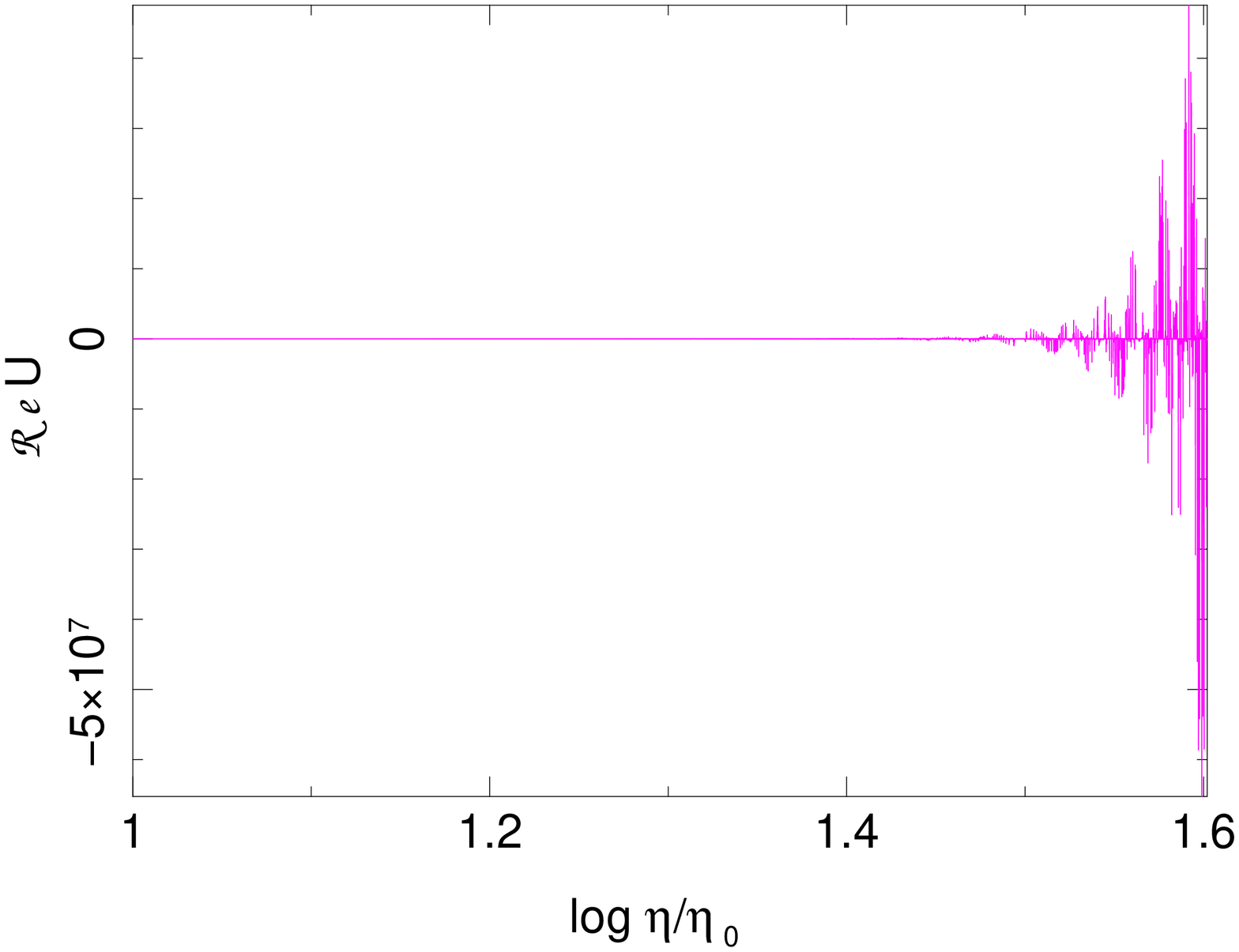} & 
\includegraphics[height=5cm,angle=-90]{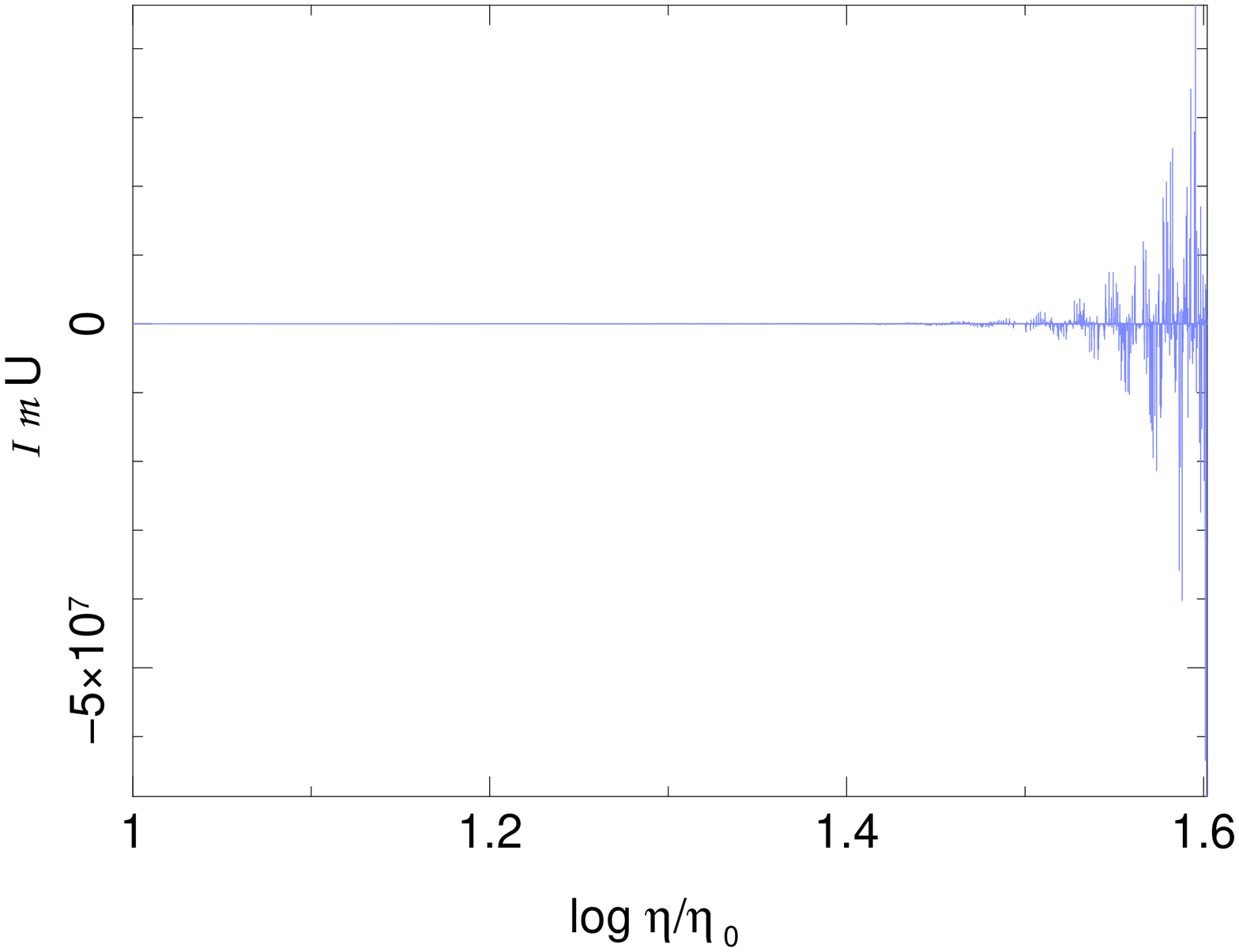} &
\includegraphics[height=5cm,angle=-90]{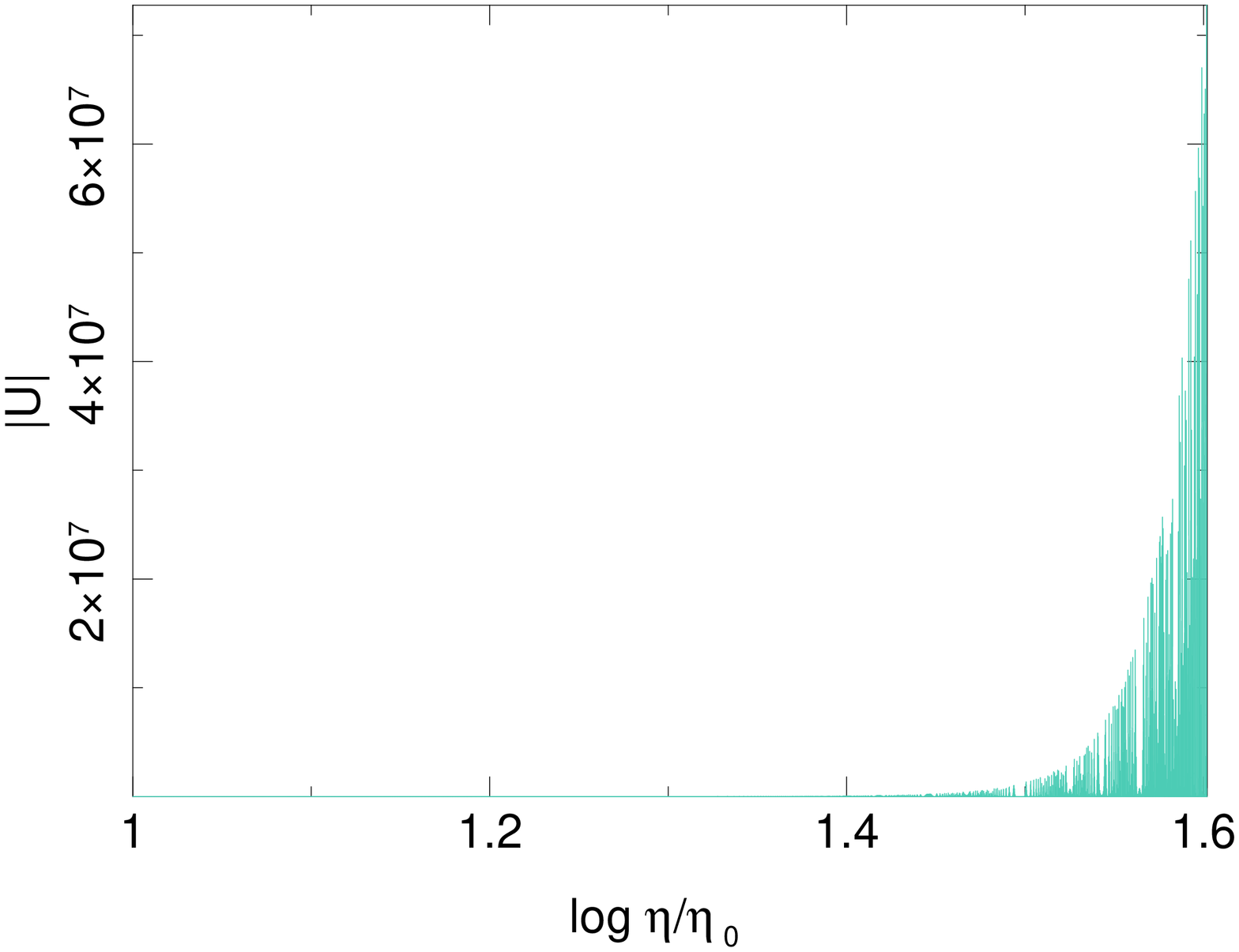} \\
\includegraphics[height=5cm,angle=-90]{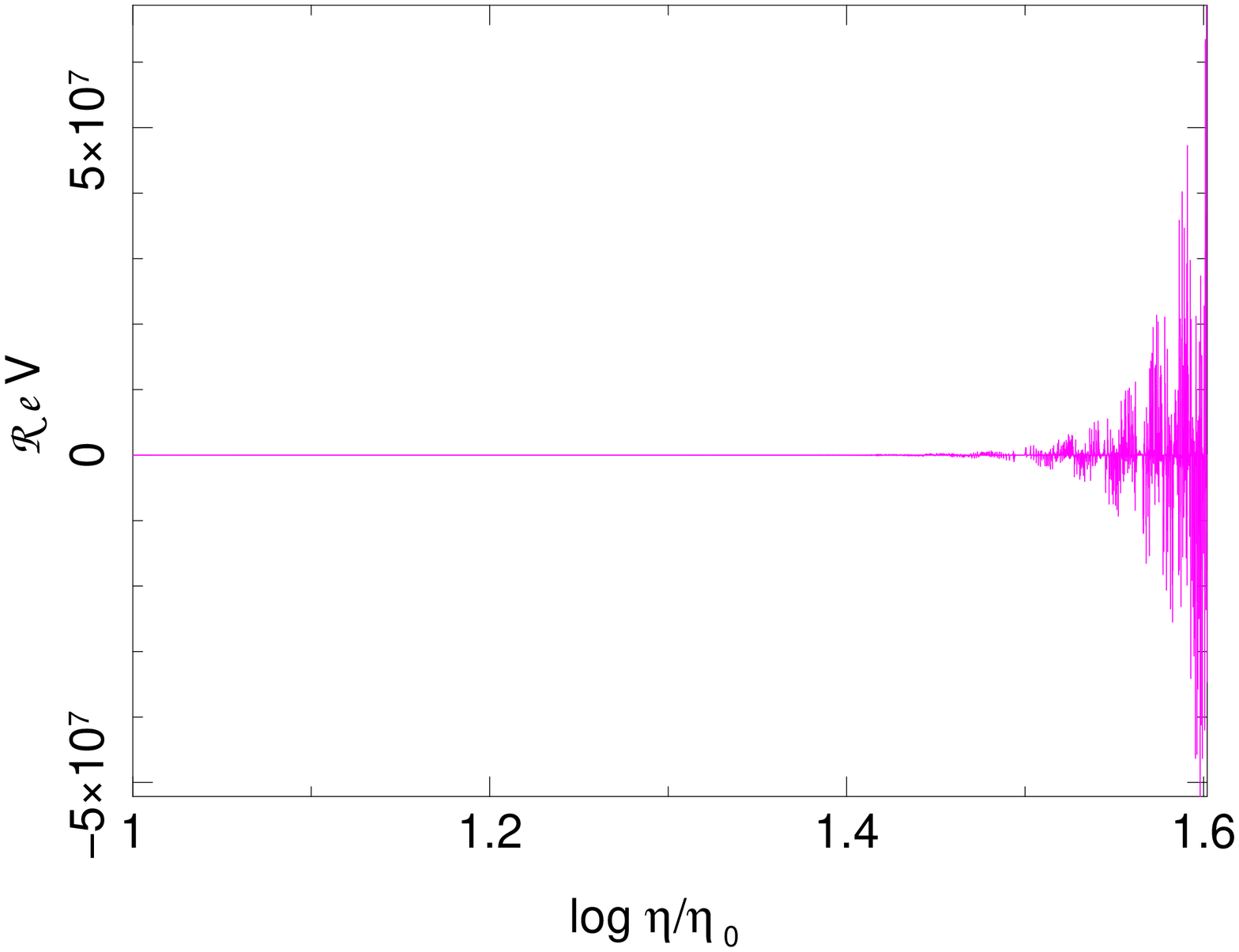} & 
\includegraphics[height=5cm,angle=-90]{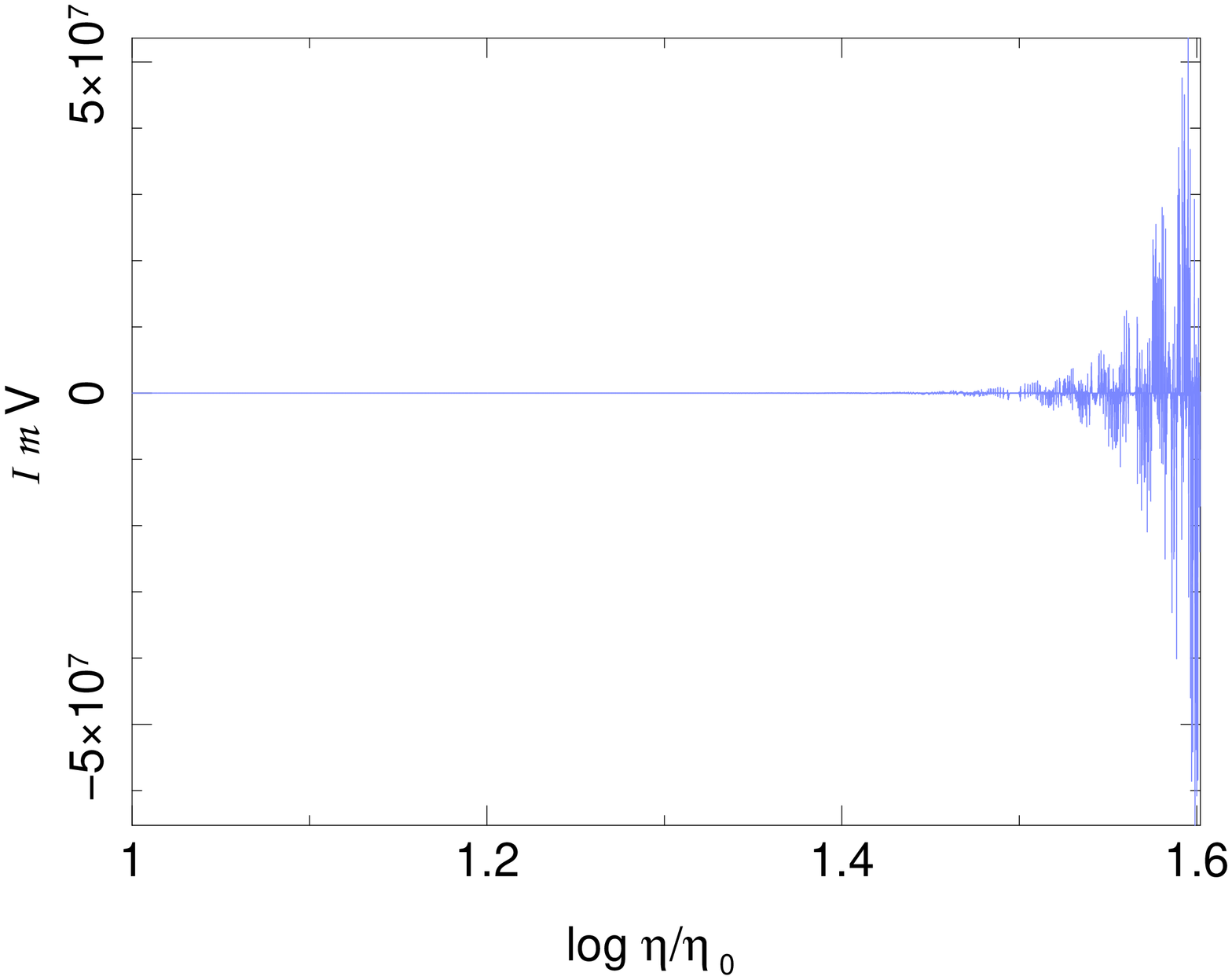} &
\includegraphics[height=5cm,angle=-90]{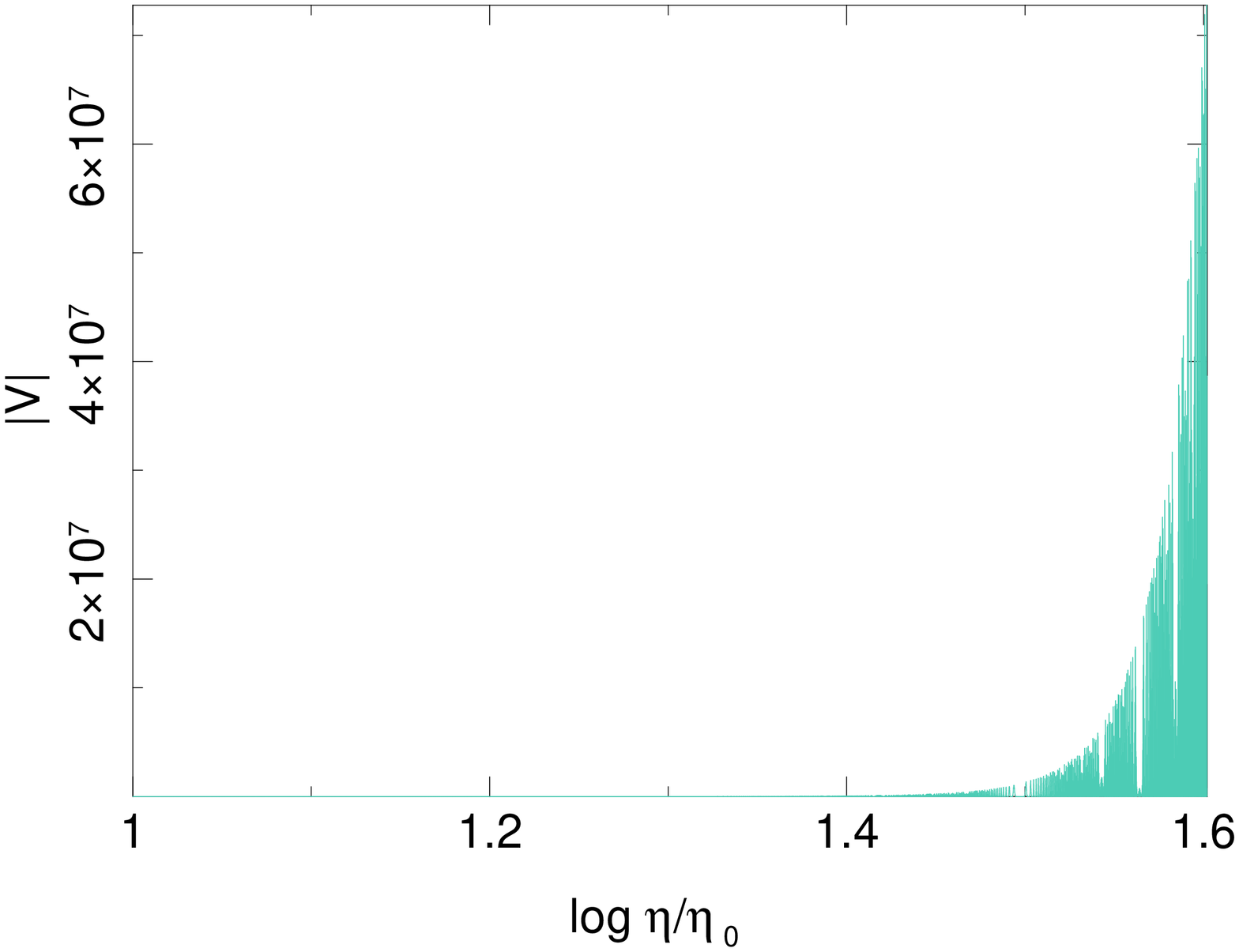} 
\end {tabular}
\end {center}
\caption{Real, imaginary, and absolute value of one of the terms respectively in $U_k$ and 
$V_k$. We used $\alpha = 0$ and $\beta = 100$. The general aspects of these functions are 
not very sensitive to $\alpha$ and are very similar for large $\beta \gtrsim 10$. Note that 
although there are resonant jumps in the value of $U$ and $V$, due to the complexity of the 
interaction term, they are not regular like in preheating case.
\label{fig:uv}}
\end{figure}

It is useful to compare this result with exponential particle production during preheating. 
In fact equation (\ref{aint}) has the same structure as the linearized equation for the 
quantum fluctuations around the minimum of the inflaton potential. But here the effective 
potential is more complex than many of inflationary models. In addition, it contains 
space-time dependent coefficients. Nonetheless, the general aspects of the asymptotic 
behaviour of these models are similar, compare Figure \ref{fig:uv} with e.g. figure 3 in
Ref.~\cite{preheat}. The reason for this similarity is the fact that exact solutions of the 
non-perturbed equation in both cases are the same, and in our case quantum corrections 
include Green's functions which are again the solution of the same type of equation with 
different boundary conditions. They are combined in a sophisticated manner - through 
integration which in the approximate solutions (\ref{uapprox}) and (\ref{vapprox}) acts like 
a linear operation.

An exponential growth of the condensate for ever would be evidently catastrophic for this 
model. We will see later that when the radiation domination ends, the faster expansion of 
the Universe during matter domination stops the growth. On the other hand, if $X$ has a short 
lifetime and the decay ends before the end of radiation domination epoch, production term 
becomes negligibly small and stop the growth. This is reflected on the approximate effective 
temperature of $A$ particles obtained in Appendix \ref{app:a}. The maximum amplitude of the 
condensate depends on the decay model, the lifetime of $X$, its coupling to $\Phi$, 
masses, cosmological parameters such as $H_0$, and reheating temperature. We leave the 
numerical estimation of the quantities to a future work in which we solve the evolution equation 
of the condensate numerically. Nonetheless, classical treatment of the similar 
models~\cite{houridmquin} shows that a condensate behaving like dark energy can be obtained 
for a large part of the parameter space without fine tuning.

The self-interaction term in (\ref{evola}) is highly non-linear:
\bea
{\mathcal G}^{(a)}(x) & \equiv & \lambda a^{4-n} \chi^{n-1}(x) - \frac{i \lambda^2 
a^{4-n}}{n} \sum_{i=0}^{n-2} (i+1) \binom{n}{i+1} {\chi}^i (x) \int \sqrt{-g} d^4y 
\chi^{i+1} (y) \nonumber \\
&&\biggl [[G^{(\Upsilon) >} (x,y)]^{n-i-1} - [G^{(\Upsilon) <} (x,y)]^{n-i-1} \biggr ] 
\label {aintphi}
\eea
Note that we have separated the term corresponding to the classical potential. The 
term proportional to $\langle \phi \rangle$ is canceled out because the latter is null by 
definition. Assuming that the coupling $\lambda$ is small, we can linearize (\ref{aintphi}) 
around $\chi_0 = \um$ which is the solution of (\ref {aint}) when both couplings are 
zero. Even after linearization equation (\ref {aintphi}) becomes a differentio-integral 
equation unless we replace all $\chi$'s inside the integrand by $\chi_0$, except for $i=0$ 
term. Similar to the production term, the linearized interaction can be considered as a 
space-time dependent correction of the mass term which has the following expression:
\bea
{\mathcal G}^{(a)}(\eta, k)a^{-2}(\eta) & \approx & \lambda a^{2-n} {\chi_0}^{n-2}(x) - 
\frac{i \lambda^2 a^{2-n}}{n} \biggl \{\int dy^4 \sqrt{-g} \biggl [[G_{\Upsilon}^> 
(x,y)]^{n-1} - [G_{\Upsilon}^< (x,y)]^{n-1} \biggr ] \biggr \} + \nonumber \\
&& \sum_{i=1}^{n-2} (i+1) \binom{n}{i+1} {\chi_0}^{i-1} (x) \int dy^4 \sqrt{-g} 
\chi_0^{i+1} (y) \biggl [[G_{\Upsilon}^> (x,y)]^{n-i-1} - [G_{\Upsilon}^< (x,y)]^{n-i-1} 
\biggr ] \biggr \} \nonumber \\
\label {aintdecomp}
\eea
Remind that according to (\ref {propagphi}) propagators of $\phi$ have a term that depend 
on $\chi$ and induces a dynamical mass for $\phi$. This affects quantum corrections and 
their amplitude decreases with growing $\varphi$. We should also replace them with $\chi_0$. 
The second complexity comes from the fact that it is non-linear in $\chi_0$ and its Fourier 
transform can not be described using $\um_k$ - it depends on auto-correlation of $\um_k$'s. 
To simplify the expression we replace correlations with a simple multiplication. This is a 
rough approximation, but if $\um_k$ is dominant at $k \sim 0$, this should not be a too bad 
approximation. Finally, after these simplifications the WKB correction term is modified to 
$\wm_k'^{(a)}$:
\bea
\wm_k'^{(a)} &\equiv & \wm_k^{(a)} + \zm_k^{(a)} \label{aintphiwkb} \\
\zm_k^{(a)} &\equiv & a^{2-n} (\eta) \int d\eta \biggl (\lambda \chi_0^{n-2} (\eta, k) - 
\frac{i \lambda^2}{n} \biggl \{\int d\eta' \sqrt{-g} \int dk_1^3 \ldots dk_{k_{n-1}}^3 
\nonumber \\
&& \biggl [G_{k_1}^{(\Upsilon) >} (\eta,\eta') \ldots G_{k_{n-1}}^{(\Upsilon) >} (\eta,\eta') - 
G_{k_1}^{(\Upsilon) <} (\eta,\eta') \ldots G_{k_{n-1}}^{(\Upsilon) <} (\eta,\eta') \biggr ] + 
\sum_{i=1}^{n-2} (i+1) \binom{n}{i+1} {\chi_0}^{i-1} (\eta, k) \nonumber \\
&& \int d\eta' \sqrt{-g} \chi_0^{i+1} (\eta', k) \int dk_1^3 \ldots dk_{k_{n-i-1}}^3 
\biggl [G_{k_1}^{(\Upsilon) >} (\eta,\eta') \ldots 
G_{k_{n-i-1}}^{(\Upsilon) >} (\eta,\eta') - \nonumber \\
&& G_{k_1}^{(\Upsilon) <} (\eta,\eta') \ldots 
G_{k_{n-i-1}}^{(\Upsilon) <} (\eta,\eta') \biggr ] \biggr \} \biggr ) \label{aintphiwkbz}
\eea
From preheating results~\cite{preheat} we know that even without taking into account quantum 
corrections, the first term in (\ref{aintphiwkb}) leads to a resonant amplification. 
However, the presence of the factor $a^{2-n}$ in front of the integral over $\eta$ in 
(\ref{aintphiwkb}) reduces the relative importance of this term at late times except when 
the self-coupling is much larger than the coupling of $\Phi$ to $X$. But, in such models a 
strong clustering of the condensate is expected, which has not been observed in dark energy. 
Therefore, this type of models are not suitable as dark energy candidate.

Quantum corrections of the self-interaction have only a minor effect on the total growth 
of the condensate in this regime, because at lowest order they depend on $\lambda^2$ rather 
than $\lambda$ in the classical term. We can make the same type of simplifying approximation 
that we have discussed above for the WKB integrals of production term and we obtain very 
similar expressions for the integrals - at least asymptotically. If we assume that the 
fraction of non-condensate $\Phi$ is negligible, $\Upsilon$ propagators (\ref{propfuphi}) and 
(\ref{proppaphi}) contain only terms proportional to $C$ and its conjugate. 
Equation (\ref{condc}) shows that $C$ also has the general form of the other expressions 
which are originated from (\ref {parbcylu}) and (\ref {parbcylv}). Therefore after 
integration, the contribution of the quantum correction of self-interaction has the same 
general behaviour as the classical term and quantum corrections due to production - 
interaction with $X$. There is however, a major difference between production and 
self-interaction. The quantum field $\phi$ has a dynamical mass that depends on the 
amplitude of the condensate. With the exponential growth 
of $\varphi$ the effective mass of $\phi$ increases and the amplitude of its propagator 
i.e. the cross-section of self-interaction decreases. This also affects 
$\langle \phi XA\rangle$ in model (c) because its closed time path integrals contain a $\phi$ 
propagator. Therefore, in this regime the self-interaction can even decreases the amplitude 
of the condensate. It can be interpreted schematically as the recombination of $\varphi$ and 
production of free $\phi$ particles.

We conclude that a classical scalar field - a condensate - can be formed during the radiation 
domination epoch with or without self-interaction. As the growth is exponential, the coupling 
constants must be very small to prevent the over production of the condensate. Here we  
should remind that in the calculation presented here we didn't couple the expansion factor 
$a (\eta)$ to the evolution of the condensate and assumed that other constituent of the 
Universe, notably relativistic particles, dominant the energy density - the reason for calling 
this epoch {\it radiation domination}. In reality the evolution of the condensate and 
expansion factor are coupled and if the condensate dominates, the expansion factor 
grows more rapidly and reduces the growth of the condensate due to the presence of negative 
power of $a (t)$ in the field equation and interactions. This fact becomes more clear in the 
next section where we study the evolution of the condensate during matter domination epoch. 

Before finishing this section we quickly review the behaviour of the two other models. Model 
(b) is very similar to model (a) but with an additional propagator in the closed time path 
integrals and an additional $a^{-1}$ factor. Each one of them adds a factor of 
$(\eta/\eta_0)^{-1/2}$ to integrals in $\wm_k^{(b)}$, the analogue of $\wm_k^{(a)}$ for this 
model. Therefore, $\wm_k^{(b)}$ has 3 propagators in closed time path integrals and an 
additional $(\eta/\eta_0)^{-1}$ factor. Although both these terms decrease with time, due to 
the exponential growth of $U_k$ and $V_k$ the condensate continues to grow exponentially but 
slower than model (a). 

In model (c) there are two expectation value due to the interaction of $\Phi$ with $X$. 
The term $\langle XA \rangle_c$ apparently looks like the similar term in model (a). But, the 
$\varphi^2$ factor in (\ref {valxa}) makes it nonlinear. For obtaining an analytical solution, 
this term must be linearized and simplifications similar to what we explained for the 
self-interaction must be applied. The term $\langle \phi XA\rangle$ becomes linear if we 
replace the $\varphi$ (or equivalently $\chi$) dependent terms in $G_{\Upsilon}$ with 
$\varphi_0$ ($\chi_0)$. As we mentioned above, the amplitude of this propagator decreases with 
the growth of $\phi$. Therefore, the condensate evolution in this model should significantly 
deviate from the other models. In particular, we expect a stronger feedback between 
$\varphi$ (equivalently $\chi$) and expansion factor $a (\eta)$. 

\subsection{Matter domination epoch} \label{sec:matdom}
In this section we consider the matter dominated epoch. We limit ourselves to the time when 
the effect of dark energy is yet negligible. In the latter case we must consider the effect 
of $\varphi (\chi)$ in the expansion of the Universe. This couples all the evolution 
equations and makes the problem insurmountably difficult. It is why we avoid this regime.

The evolution equation of $\chi_k$ for model (a) in this era takes the following form:
\bea
&& {\chi}'' + (k^2 + a^2 m_{\Phi}^2 - \frac{2}{\eta^2}) \chi + \frac{i \mg^2 a \chi (k)}
{(2\pi)^6} \int d^3 k_1 d^3 k_2 \delta^{(3)} (\vec{k} - \vec{k}_1 - \vec{k}_2) \nonumber \\
&& \quad\quad \int d\eta'\sqrt{-g} \biggl [G_{k_1}^{\ca >} (\eta,\eta') G_{k_2}^{\cx >} 
(\eta,\eta') - G_{k_1}^{\ca <} (\eta,\eta') G_{k_2}^{\cx <} (\eta,\eta') \biggr ] + 
\text{self-interaction} = 0 
\label {aintmat}
\eea
This equation has an additional term due to a nonzero $\ddot {a}(\eta)$, and the expansion 
factor evolves according to (\ref{amatter}) in Appendix \ref{app:c}. It increases faster than 
radiation domination era, as $\eta^2/\eta_0^2$ than rather than linearly. Similar to 
the previous section, we first 
neglect self-interaction. The solution of equation (\ref{aintmat}) without the interaction 
term is also explained in Appendix \ref{app:c}. It has exact solutions for the special cases 
when $m=0$ or $k=0$. In Appendix \ref{app:c} we use WKB approximation for the case of 
$m \neq 0$ and $k \gtrsim 0$. Because the interaction term in (\ref{aintmat}) is 
proportional to $\chi$ we can consider it as a time-dependent mass and use again the WKB 
technique to obtain an approximate solution:
\bea
\chi_k^{(a)} (\eta) & \approx & \sqrt{\frac{\eta}{\eta_0}} \biggl \{{c'}_k^{(a)} 
J_{\frac{1}{2}} 
\biggl (\beta' \frac{\eta^3}{\eta_0^3} (1- \frac{3k^2\eta_0}{2m_\Phi^2 \eta}) + 
\ym_k (\eta)\biggr ) + {d'}_k^{(a)} J_{-\frac{1}{2}} \biggl (\beta' \frac{\eta^3}{\eta_0^3} 
(1- \frac{3k^2\eta_0}{2m_\Phi^2 \eta}) + \ym_k (\eta) \biggr )\biggr \} \nonumber \\
\label{amodelsol} \\
\ym_k^{(a)} (\eta) & \equiv & \frac{3i \mg^2 \beta'}{2 (2\pi)^3 m_\Phi^2} \int d(\frac{\eta}
{\eta_0}) \int d^3 k_1 d^3 k_2 \delta^{(3)} (\vec{k} - \vec{k}_1 - \vec{k}_2) \nonumber \\ 
&& \int d\eta'\sqrt{-g} \biggl [G_{k_1}^{\ca >} (\eta,\eta') G_{k_2}^{\cx >} 
(\eta,\eta') - G_{k_1}^{\ca <} (\eta,\eta') G_{k_2}^{\cx <} (\eta,\eta') \biggr ]
\label{wkvsol}
\eea
As usual the integrals in (\ref{wkvsol}) are very complex and long. Therefore, in place of 
presenting all the details we only discuss the general form of their terms, and their 
asymptotic behaviour when $\eta/\eta_0 \gg 1$.

At late times we both $A$ and $X$ particles are non-relativistic and their temperatures can 
be considered to be zero. This makes the expression of propagators (\ref{propfuphigen}) 
and (\ref{proppaphigen}) much simpler. As we are only interested in the asymptotic behaviour 
of the condensate, we only consider the terms with the highest order of $\eta/\eta_0$. 
In the integral over $\eta'$ the dominant terms have the following forms:
\bea
&&\int d(\frac{\eta'}{\eta_0}) \frac{\eta'^6}{\eta_0^6} \sin \alpha \frac{\eta'^3}{\eta_0^3} = 
\frac{i\alpha^{-\frac{7}{3}}}{6} [e^{-\frac{7\pi i}{6}} \gamma (\frac{7}{3}, i\alpha 
\frac{\eta'^3}
{\eta_0^3}) - e^{\frac{7\pi i}{6}} \gamma (\frac{7}{3}, -i\alpha \frac{\eta'^3}{\eta_0^3})] 
\label{intsin} \\
&&\int d(\frac{\eta'}{\eta_0}) \frac{\eta'^6}{\eta_0^6} \cos \alpha \frac{\eta'^3}{\eta_0^3} = 
\frac{\alpha^{-\frac{7}{3}}}{6} [e^{-\frac{7\pi i}{6}} \gamma (\frac{7}{3}, i\alpha 
\frac{\eta'^3}
{\eta_0^3}) + e^{\frac{7\pi i}{6}} \gamma (\frac{7}{3}, -i\alpha \frac{\eta'^3}{\eta_0^3})] 
\label{intsin}
\eea
with $\alpha = \pm \beta'_A \pm \beta'_X$. Then the dominant terms in the integral over $\eta$ 
become:
\be
\int d(\frac{\eta}{\eta_0}) \gamma (\frac{7}{3}, i\alpha \frac{\eta'^3}{\eta_0^3}) \sin (\beta 
\frac{\eta'^3}{\eta_0^3}), \quad \int d(\frac{\eta}{\eta_0}) \gamma (\frac{7}{3}, i\alpha 
\frac{\eta'^3}{\eta_0^3}) \cos (\beta \frac{\eta'^3}{\eta_0^3}) \label{intetamatt}
\ee
where $\beta = \pm \beta'_A \pm \beta'_X$. These integrals can be calculated analytically when 
the $\gamma$ function is replaced by its asymptotic expansion 
$\gamma (7/3, i\alpha \eta'^3/\eta_0^3) \approx \Gamma (7/3) - (i\alpha \eta'^3/\eta_0^3)^{4/3} 
exp (-i\alpha \eta'^3/\eta_0^3)$. Finally the approximate expression for $\chi_k$ at late 
times and without considering self-interaction is obtained as the following:
\bea
\chi_k (\eta) & \xrightarrow[\lambda = 0]{\frac{\eta}{\eta_0} \gg 1} & \sqrt {\frac{2}
{\pi \beta'_\Phi}} \frac{\eta_0}{\eta} \biggl (1 - \frac{3k^2\eta_0}{2m_\Phi^2 \eta} + 
\ym_k (\eta) \biggr ) \biggl \{{c'}_k^{(a)} \sin \biggl (\beta' \frac{\eta^3}{\eta_0^3} 
(1- \frac{3k^2\eta_0}{2m_\Phi^2 \eta}) + \ym_k (\eta)\biggr ) + \nonumber \\
&& {d'}_k^{(a)} \cos \biggl (\beta' \frac{\eta^3}{\eta_0^3} (1 - \frac{3k^2\eta_0}
{2m_\Phi^2 \eta}) + \ym_k (\eta) \biggr )\biggr \} \label{chimatapp} \\
\ym_k (\eta) & = & \frac{i \mg^2}{4 (2\pi)^3 \pi \sqrt{\beta'_A \beta'_X}} 
\biggl \{\sum_{\alpha} C_\alpha (k, \bar{x}) \gamma (-2, i\alpha \frac{\eta^3}{\eta_0^3}) + 
\nonumber \\
&& \sum_{\alpha} C'_\alpha (k, \bar{x})\gamma (-\frac{1}{3}, i\alpha \frac{\eta^3}{\eta_0^3}) 
\biggr \} \label{vmsol}
\eea
where $C_\alpha$ and $C'_\alpha$ are proportional to the distributions of $A$ and $X$ 
particles in the same way as what was described in Sec. \ref{sec:raddom}. Because in this 
regime both these particles are considered to be non-relativistic, equation (\ref{xdens}) 
is applied as their distribution. At late times 
$\gamma$ functions in (\ref{vmsol}) approach a constant and $\bar{x}$ dependent terms i.e 
terms containing $f^{(A)}$ and $f^{(X)}$ decay very rapidly, as $(\eta_0/\eta)^6$ for terms 
containing one $f$. Therefore, at late times $\chi_k (\eta)$ is an oscillating function and 
its amplitude decreases as $\eta_0/\eta$ with time. Consequently, $\varphi_k$ decreases as 
$\eta^3_0/\eta^3$ and we conclude that in the matter domination epoch where the expansion 
of the Universe is faster, the production of $\Phi$ in the decay of $X$ alone is not enough 
to compensate the expansion, and the density of the condensate will decrease. Evidently, the 
validity of this conclusion depends on how precise are the approximation considered here. 
Note also that (\ref{chimatapp}) is written for small $|k|$ and is not valid for 
$|k| \rightarrow \infty$. For the latter case the solution (\ref{solkmmatter}) for $m=0$ 
must be used. It contains also only oscillatory terms.

Equation (\ref {aintmat}) becomes non-linear if we consider also the self-interaction. 
Therefore, to obtain an analytical solution we must linearize it by replacing the interaction 
term with its linearized version the equation (\ref{aintdecomp}), and in (\ref {chimatapp}) 
$\ym_k \rightarrow \ym'_k \equiv \ym_k + \zm'_k$, where $\zm'_k$ is the same expression as 
$\zm_k$ in (\ref{aintphiwkbz}) written for matter domination epoch. 
After taking integrals, $\zm'_k$ will include terms very similar to what we 
obtained for $\ym_k$. This means that the solution of complete linearized equation also 
contains only oscillating terms, and therefore $\varphi$ decays as $\eta_0^2/\eta^2$ or 
equivalently $t_0/t$. Nonetheless, for the reasons we explain now this approach is not 
realistic. In fact when self-interaction is added to equation (\ref{aintmat}), it becomes 
a non-linear differentio-integral equation, because the propagators of $\Phi$ depend 
on the condensate. Therefore, only a solution without linearization can give a realistic 
answer to this problem. This is possible only numerically and we leave it to a future work. 
In the rest of this section we estimation the late time behaviour of condensate qualitatively.

To perform a non-linear analysis of equation (\ref{aintmat}) with self-interaction we first 
neglect quantum corrections. This means that we only consider the first term of 
(\ref {aintphi}) for which the minimum of the potential is at origin. If for simplicity we 
neglect also the production term the evolution equation becomes:
\be
{\chi}'' + (k^2 + a^2 m_{\Phi}^2 - \frac{2}{\eta^2}) \chi + \lambda a^{4-n} \chi^{n-1}(x) = 0
\label{aintclass}
\ee 
Using a difference approximation for derivatives but without linearization, we find that 
although at the beginning $\chi$ can grow, whatever the initial conditions, at late times 
it approaches to zero. This means that this equation lacks a tracking solution. Another way 
of checking the absence of a tracking solution is the application of the criteria 
$\Gamma \equiv V"V/V'2 > 1$ suggested as the necessary condition for the existence of a 
tracking solution~\cite{trackingcond}. For equation (\ref{aintclass}) 
$\Gamma = n (n-1) / n^2 < 1$ for $n > 0$. This is a well known result that only inverse 
power-law and inverse exponential potentials have a late time tracking solution~\cite{quin}.

The situation is different when we add quantum corrections. The coefficient $C$ in 
(\ref{propfuphi}) and (\ref{proppaphi}) which determines the amplitude of the quantum state 
of the condensate depends inversely on $\chi$. In another word, there is a back reaction 
from the formation of the condensate on the propagators of $\phi$. 
To estimate the time evolution of the solutions of equation (\ref{intetamatt}) we use the 
approximate solution of field equation (\ref{homosolmatter}) and determine the linearized 
self-interaction term (\ref{aintdecomp}). Similar to simplifications applied to the 
evolution equation in radiation domination case, we consider $\bar {x}$ as an external 
parameter and when we want to solve (\ref{intetamatt}), we neglect its space dependence and 
identify $\bar{\eta}$ with $\eta$. The counting of $\chi$ exponents in (\ref{aintphi}) shows 
that due to the presence of negative power of $\chi$ in (\ref{condc}) some of the terms in 
the quantum correction, will be proportional to a negative power of $\chi$. This means that 
quantum corrections play the role of an inverse power-law potential with varying 
coefficients. As mentioned above, this type of potential is one of the well studied 
candidates for quintessence models and have a tracking solution~\cite{quin}. The only 
difference here is the time dependence of coefficients. Therefore, we must find the 
conditions under which these coefficient are constant or vary slowly.

It is easy to verify that the term with $i=0$ in (\ref{aintphi}) includes the highest 
negative power of $\chi$. Therefore, it dominantly determines the tracking behaviour of the 
field. Fortunately, it has also the simplest integral in the sense that when we assume 
$\bar {x}$ as an independent variable and integrate over $\eta'$, the integral contains a 
linear power of $\chi$. After taking the Fourier transform, $\chi_k$ is integrated out and 
the integrand contains only the propagators. For $i > 0$ terms the integrand includes also 
$\chi$ factors and the evolution equations (\ref{evola})- (\ref{evolc}) become 
differentio-integral equation. For simplicity here we neglect these subdominant terms. 
After integration the evolution equation of the condensate in matter domination epoch becomes:
\bea
&& {\chi}'' + (k^2 + a^2 m_{\Phi}^2 - \frac{2}{\eta^2}) \chi + \frac{i}{3}\lambda^2 
a^{4-n} (\frac{2}{\pi \beta'_\Phi})^{n-2} e^{i\frac{(8-n)\pi}{6}} (\frac{\eta_0}{\eta})^{n-1} 
\sum_{\alpha,\beta} \beta^{-\frac{8-n}{3}} \gamma (\frac{8-n}{3}, -i\beta 
\frac{\eta^3}{\eta_0^3})~ e^{i (\alpha + \beta) \frac{\eta^3}{\eta_0^3}} 
\nonumber \\
&& \quad \quad \times \sum_{i = 1}^{n-1} \binom{n-1}{i} (\frac{2}{\pi\beta'_\Phi})^{n-i} 
\cos^{2(n-i)}(\beta'_\Phi \frac{\eta^3}{\eta_0^3})~(\frac{\eta_0}{\eta})^{2(n-i)}~
\chi^{-2(n-i)+1} (\eta) +  \ldots = 0 \nonumber \\
&& \alpha, \beta = j\beta'_\Phi, \quad \quad j = -(n-1), 
\ldots, n-1 \label{chiselfevol}
\eea
where dots indicates subdominant terms, including the production term. Therefore 
equation (\ref{chiselfevol}) presents the dominant terms of the evolution equation for 
all three decay modes considered in this work. Ignoring the time-dependence of coefficients,  
it is evident that the effective potential in (\ref{chiselfevol}) satisfies tracking 
condition explained above because it is a polynomial of inverse powers of $\chi$. But there is 
no known existence condition for tracking when coefficients are time dependent. The best guess 
is that if some of the coefficients approach a constant or vary slowly, the solution can be 
roughly a tracker. In fact we notice that due to negative powers of $\eta/\eta_0$ in 
(\ref{chiselfevol}) many of coefficients become negligibly small asymptotically. 
By counting the order of $\eta/\eta_0$ terms, using the asymptotic expression of 
incomplete $\gamma$ function, we conclude terms satisfying the following conditions have 
slowest variation and are constant or growing when $\eta/\eta_0 \gg 1$:
\be
\alpha = - 2\beta, \quad\quad 17 - 6n + 2i \geqslant 0 \label {powercount} 
\ee
The first condition eliminates the oscillatory factors, and the second one is the order 
of $\eta/\eta_0$ 
factors. As $i \leqslant n-1$, this condition is satisfied only for $n \leqslant 3$. This is 
the only model in which the condensate does not decay quickly and a tracking solution does 
possibly exist. The case of $n=4$ is also interesting because although apriori it does not 
have positive index term, it includes terms that decay only linearly with time, and 
therefore can lead to solutions in which the condensate decays with $w > -1$, but enough 
slowly to be consistent with data. We should also remind that renormalization of the complete 
theory in general induces anomalous dimensions for the fields that somehow modifies the 
exponents of the bare theory. In addition, although the production term decays with time, 
at intermediate epochs where matter is yet the dominant component its effect can be 
significant. 

Considering all these uncertainties, we conservatively conclude that for small $n$'s such 
as $n=3$ and $4$ the decay models considered here seem to produce a condensate that grows 
exponential in the radiation domination epoch and asymptotically evolves to a constant 
field in matter domination era. We note that these values for the self-interaction order 
are the only renormalizable polynomial potentials in 4-dimension space-times. When the 
energy density of the condensate 
becomes dominant the study of its behaviour is more complex because the evolution equations 
of the condensate and expansion factor $a (\eta)$ become strongly coupled. Nonetheless the 
classical treatment of the same type of models in~\cite{houridmquin} showed that the 
tracking behaviour persists in this regime, at least until the matter contribution in 
the expansion is not completely negligible. 

The amplitude of the effective interaction term in (\ref{chiselfevol}) depends inversely on 
mass. Therefore fields with large mass produces smaller condensate. This is consistent with 
our initial assumptions that heavy scalars $X$ and $A$ don't condensate. Moreover, we note 
that $\beta'_\Phi \gg 1$. This means that even for relatively large $\lambda$ the asymptotic 
value of $\varphi$ can be small. This increase the initial probability of the formation of 
condensate and at same time reduces this amplitude at late times and reduces the 
fine-tuning of couplings.

\subsection{Initial conditions for the condensate} \label{sec:initcondcon}
To complete the study of the evolution of condensate we must fix the initial conditions. 
It is specially important for the asymptotic value of the condensate which determines its 
density today, and therefore constrains the mass and couplings of $\Phi$. As we do not have 
an explicit expression for the solution of (\ref{chiselfevol}) the initial conditions do not 
add any information to what we have obtained so far. Therefore, the discussion of this section 
is for the sake of completeness and would be useful future numerical simulation of this 
model.

We assume that before the decay of $X$ particles there is no condensate or more strongly 
there is no $\Phi$ particle. In this case the initial condition for $\chi$ is trivial, 
$\chi (\eta = \eta_0) = 0$. But $\chi' (\eta = \eta_0)$ in general should be non-zero and 
in fact positive. Its value presents the initial production rate of the condensate and is 
related to the initial density and the decay rate of $X$ particles. 

It is expected that the Boltzmann equations which determine the evolution of distributions 
depend on $\varphi$ but not on its derivatives, see Appendix \ref{app:a}. Moreover, they are 
first order differential equations and each of them needs only one initial condition which 
can be chosen to be the initial distribution of the species. If we solve the evolution 
equation of $\varphi$ along with the Boltzmann equations for all the species, this single set 
of inputs is enough to completely solve the Boltzmann and the condensate equation. In fact, 
once a solution of the Boltzmann equations as a functional of $\varphi$ is found, by taking 
its derivative one can determine the initial value of $\varphi'$. This operation provides the 
complete initial conditions for the condensate's evolution equation. Here however, we do not 
solve Boltzmann equations in details. Thus, we can not follow this procedure and must guess 
the initial value of $\varphi'$ based on some physical properties. We remind that Boltzmann 
equations present the conservation of the flux of particles in the phase space. Thus we use 
the particle number conservation along with the additional assumption that initially, 
all the $\Phi$ particles join the condensate immediately. We also neglect their spatial 
anisotropy. Under this conditions we obtain the following condition for the initial value 
of $\chi'$:
\be
\chi' (\eta_0) \approx \biggl (a(\eta_0) n^{(X)} (\eta_0) \eta_0 \Gamma m 
{\mathcal M}_\Phi \biggr )^{\frac{1}{2}} \label{initchip}
\ee
Another way of determining integration constants is to fix the initial and final value 
of $\chi$. However, for the reasons explained above it will over-constrain the model when 
the full model is solved. For a tracking solution of (\ref{chiselfevol}) the initial 
conditions don't affect the late behaviour of the condensate but they determine its 
asymptotic value.

\section{Discussion}\label{sec:conclu}
Although we have not yet observed any elementary scalar field, they are believed to play 
important roles in the foundation of fundamental forces and phenomena that have shaped our 
Universe. On the other hand, we have observed the composite scalar fields and their 
condensation in condense matter where they are responsible for various phenomena. such as 
symmetry breaking, mass acquisition of photons, quantization of flux tubes, formation of 
topological defects, and many other exotic behaviours of matter. From these findings we have 
learned that the condensate has usually a simple potential which can be related to the 
interactions in the original Lagrangian - at least qualitatively. For instance, in the case 
of Cooper pairs in superconductors, the presence of a $\varphi^4$ potential can be 
schematically interpreted as an elastic scattering between electrons inside two Cooper pairs 
due to electromagnetic interaction. At low energies it is seen as a point-like scattering of 
2 incoming scalar particles to two outgoing scalars - similar to self-interaction of a scalar 
field. 

The potential of the condensate of a fundamental scalar field should also trace back to 
the Lagrangian of the quantum field. Many particle physics applications of scalar 
condensates only consider the classical order which correspond to the potential in the 
Lagrangian. Because Lagrangians are usually local, the effect of the classical term is also 
local. However, as we have demonstrated in the previous sections, in some circumstances the 
effect of quantum corrections can be very crucial. In fact the properties of the condensate 
discussed in the previous sections, and in Appendices \ref{app:a} and \ref{app:b} are related 
to the non-local properties of quantum mechanics, and by extension quantum field theory. 
The descriptions of the quantum state of a condensate suggested in Ref.~\cite{condwave} 
and its generalization in Appendix \ref{app:b}, include infinite number of entangled 
particles. Non-locality of quantum mechanics assures that these particles {\it feel} the 
presence of each others even at largest cosmological distances, and therefore behave 
collectively at large scales. At early times when the Universe is dense and the probability 
of scattering between particles is large, the local effect of the classical potential 
as well as the production 
of $\Phi$ particles by the decay of $X$ particles, which at lowest order is like a classical 
scattering, controls the amplitude of the condensate and the distribution of free $\phi$
particles. But due to the expansion of the Universe, the cross-section of interaction and 
scattering becomes smaller at late times, and non-local effects become dominant. The very 
small coupling of $\Phi$ with other fields assures the stability of the condensate. It could 
be destroyed if interactions were strong enough to wash out the coherence of the condensate 
state, i.e. made particles to behave individually and semi-classically. 

Apriori it should be possible to design experiments or observations capable of distinguishing 
between a condensate of quantum origin and a fundamentally classical field as dark energy. 
If the latter case is true, dark energy must be due to a modification of the general 
relativity which is believed to be a classical theory at least for scales 
$k \lesssim M_{Planck}$. In this case the classical field would have a purely geometrical 
origin. On the other hand, if dark energy is produced by a condensate we expect to see some 
quantum effects. For instance to be able to observe its quantum excitation, similar to the 
excitation of a Bose-Einstein condensate~\cite{bec}. Evidently, due to the extremely small 
coupling of dark energy, the production of such excitations in the lab or their observation 
in cosmological environment is extremely difficult, if not impossible. Nonetheless, with the 
progress of our understandings in condense matter physics and related technological advance, 
there is hope that one day such an exploration become possible.

If numerical simulations confirms our conclusions about the order of self-interaction 
potential, this would be a very significant result. In quantum field theories the dimension 
of an interaction term determines its renormalizability. For scalar fields in 4-dimension 
spacetimes $\Phi^n$ with $n \leqslant 4$ are renormalizable. In fact except this physically 
motivated requirement there is no other rule to constrain the self-interaction in a quantum 
field theory. Interestingly enough, these values for $n$ correspond exactly to models for 
which a late time cosmological condensate seems to excite. In the $n=3$ models the coupling 
has a mass dimension of one which is considered to be the vacuum expectation value of another 
field, or to be proportional to the Planck mass, the only dimensional fundamental constant 
we know. In the latter case the field $\Phi$ is probably related to quantum gravity models. 
In addition, it must be a singlet or a 1-form in the group manifold of the symmetry group 
of the model, otherwise it breaks the symmetry. On other hand, in a $n=4$ model the coupling 
constant is dimensionless and $\Phi$ can be in a nontrivial self-conjugate representation of 
the symmetry group. These observations help to constrain the candidate particle physics 
models of dark energy. We can also put a rough lower limit on the self-coupling of 
$\Phi$. $\Gamma_\Phi$ the interaction width of $\Phi$ must be larger than $H_0$ if $\Phi$ is 
not yet freezed-out. For $n=4$ self-interaction this means 
$\Lambda_\Phi \gtrsim \sqrt{H_0 m_\Phi}$, and for $n=3$ potential 
$\Lambda_\Phi/M_{Planck} \gtrsim \sqrt{H_0 m_\Phi}/M_{Planck}$~\cite{houridmquin}. 

In summary, we used quantum field theory techniques to study the condensation of a scalar 
field produced during the decay of a much heavier particle in a cosmological context. Such 
a process had necessarily happened during the reheating of the Universe. It can also happen 
at later times if the remnants of the decay don't significantly perturb primordial 
nucleosynthesis. We showed that one of the necessary conditions for the formation of a 
condensate is its light mass. By considering three decay models and a power-law 
self-interaction potential of arbitrary positive order, we showed that the self-interaction 
has an important role in the cosmological evolution of the condensate and its 
contribution to dark energy. In particular, we showed that only a self-interaction of order 
3 or 4 can produce a stable condensate in matter domination epoch. These results are obtained 
analytically and by considering a number of simplifying approximations. Therefore, they need 
confirmation by a more precise calculation, which in the face of complexity of this model, 
must be numerical. With little modification or adaptation, most of the formulation and results 
of this work are applicable to other cosmological phenomena which are based on a scalar 
condensate. 

{\bf We finish this section by reminding that if dark energy is the condensate of a scalar 
field, the importance of the quantum corrections in its formation and its behaviour found 
here is the proof of the reign of Quantum Mechanics and its rules at largest observable 
scales.} 

\vspace{1.cm}
\begin {appendix}
{\Large \bf Appendixes}
\section{Free field Green's function on non-vacuum states} \label{app:a}
In canonical representation, a free scalar field $\Phi$ can be decomposed to 
creation and annihilation operators on the Fock space:
\be
\Upsilon \equiv a (\eta)\Phi (x) = \sum_k [\um_k (x) a_k + \um_k^* (x) a^{\dagger}_k] \quad , 
\quad [a_k, a^{\dagger}_{k'}] = \delta_{kk'} \quad [a_k,a_{k'}] = 0 \quad 
[a^{\dagger}_k,a^{\dagger}_{k'}] = 0\label{canon}
\ee
where $\um_k (x) \equiv \um_k (\eta)e^{-i\vec{k}.\vec{x}} $ is a solution of 
the free field equation (\ref{eqhomo}). The quantization of $\Phi$ imposes the 
following relation on $\um_k (x)$:
\be
\um^{'}(x) \um^*(y) - \um (x)\um^{'*}(y) = \frac {-i}{a} 
\delta^{(4)} (x-y) \label{quantcond}
\ee
A Fock state $|\Psi\rangle$ is constructed by multiple application of the 
creation operator $a^{\dagger}_k$ on the vacuum state $|0\rangle$:
\bea
&& a_k |0\rangle = 0,~ \forall k \quad , \quad |k_1 k_2 \ldots 
k_n\rangle 
\equiv a^\dagger_{k_1} a^\dagger_{k_2} \ldots a^\dagger_{k_n} |0\rangle 
\label{vacdef} \\
&& |\Psi\rangle = \sum_{k_1 k_2 \ldots k_n} \Psi_{k_1 k_2 \ldots k_n} 
|k_1 k_2 \ldots k_n\rangle 
\label{psistate}
\eea
The 2-point free Green's function of $\Phi$, it can be written as:
\bea
iG_F (x,y) &\equiv &\langle\Psi|T\Phi(x)\Phi(y)|\Psi\rangle = \nonumber \\
&& \sum_k \sum_i \sum_{k_1 k_2 \ldots k_n} \delta_{kk_i} |\Psi_{k_1 k_2 
\ldots k_n}|^2 \biggl [\um_k^* (x)\um_k (y) \Theta (x_0 - y_0) + 
\um_k (x)\um_k^* (y) \Theta (y_0 - x_0)\biggr ] + \nonumber \\
&& \sum_k  \biggl [\um_k (x)\um_k^* (y) \Theta (x_0 - y_0) + \um_k^* (x)
\um_k (y) \Theta (y_0 - x_0)\biggr ] \label{propst}
\eea
From (\ref{propst}) we can extract the expression for advanced and retarded propagators:
\bea
iG^> (x,y) &\equiv& \langle\Psi|\Phi(x)\Phi(y)|\Psi\rangle = \sum_k \sum_i 
\sum_{k_1 k_2 \ldots k_n} \delta_{kk_i} |\Psi_{k_1 k_2 \ldots k_n}|^2 
\um_k^* (x)\um_k (y) + \nonumber \\
&& \sum_k \biggl [1 + \sum_i \sum_{k_1 k_2 \ldots k_n} \delta_{kk_i} 
|\Psi_{k_1 k_2 \ldots k_n}|^2 \biggr ] \um_k (x)\um_k^* (y) \label{propfu} \\
iG^< (x,y) &\equiv& \langle\Psi|\Phi(x)\Phi(y)|\Psi\rangle = \sum_k \sum_i 
\sum_{k_1 k_2 
\ldots k_n} \delta_{kk_i} |\Psi_{k_1 k_2 \ldots k_n}|^2 \um_k (x)
\um_k^* (y) + \nonumber \\
&& \sum_k \biggl [1 + \sum_i \sum_{k_1 k_2 \ldots k_n} \delta_{kk_i} 
|\Psi_{k_1 k_2 \ldots k_n}|^2 \biggr ] \um_k^* (x)\um_k (y) \label{proppa} 
\eea
Therefore, for a free fields the Feynman propagator $G_F (x,y)$ on a non-vacuum state can 
be written as a linear expansion with respect to the independent solutions of the free 
field equation $\um_k (x)\um_k^* (y)$ and $\um_k^* (x)\um_k (y)$. The contribution 
of a non-vacuum state $|\Psi\rangle$ appears in the coefficients of the expansion. 

When the entanglement and interaction between particles are negligible, the N-particle 
wave function $|\Psi\rangle$ can be written as a direct product of the 1-particle states. 
In this case, the projection coefficients are $|\Psi|^2 \approx f (k, \bar {x})$, where 
$f (k, \bar {x})$ is the classical energy-momentum and space-time distribution of 
particles. In fact, if we project $\Psi$ to the coordinate space we can express $|\Psi|^2$ 
as a functional of Wigner function~\cite{wigner}:
\be
|\Psi|^2 = \Psi^*(x) \Psi(y) = \Psi^* (\bar{x}+\frac{X}{2}) \Psi (\bar{x}-\frac{X}{2}) = 
\frac{\sqrt {-g}}{(2\pi)^4} \int d^4 p P(\bar{x},p) e^{-ip.X} \quad , \quad \bar{x} \equiv 
\frac {x+y}{2} \quad , \quad X \equiv x-y
\ee
In the classical limit the Wigner function $P(\bar{x},p)$ approaches the classical 
distribution function $f (p, \bar {x})$. As explained in Sec. \ref{sec:prog}, these 
distributions can be determined in a consistent way from the classical Boltzmann equations 
(\ref{bolt}) or their quantum extensions Kadanoff-Baym equations~\cite{noneqqft,nonequi}. 
However, Boltzmann equations of interacting species are coupled to each others and to the 
evolution equation of the condensate. Thus, a complete solution can be obtained only by 
numerical calculation. Nonetheless, we need the distribution of particles to be able to 
to even an approximate solution for the condensate. For this reason we simplify the problem 
by assuming that $f^{(i)} (p, \bar {x})$ for $i=X,~A,~\phi$ have thermal distribution and 
the effect of interactions can be included in the variation of their temperature.

The number density of $X$ particles decreases by a factor of $\exp (-(t-t_0)/\tau)$ where 
$t_0$ is an initial time. We assume that $X$ particles are non-relativistic since their 
production i.e. $T_X \ll m_X$. Note that the decay is a non-thermal process. But when it 
happens slowly, the deviation from a thermal distribution is small and can be approximated 
by an effective time-dependent temperature. As $X$ is non-relativistic its number density is 
approximately proportional to $T_X^{3/2}$. This means that due to the decay $T_X$ decreases 
by a factor proportional to power $2/3$ of the decay term. We must also take into account 
the effect of the expansion of the Universe. Therefore, under these approximations: 
\be
f^{(X)} (p, \bar{x}) \approx \frac{1}{e^{p^\mu \beta_\mu} - 1} \quad, \quad |\beta|^{-1}
\equiv {\mathsf k}_B T^{(X)} (\bar {x}) \approx \frac {{\mathsf k}_B 
T^{(X)}(a_0) D^{\frac{2}{3}}(\bar{x}) a^2 (t_0)}{a^2 (t)} \exp (- \frac{2 (t-t_0)}
{3 \tau})  \label{distx}
\ee
where $D (\bar{x})$ is the growth factor of fluctuations, ${\mathsf k}_B$ is the 
Boltzmann constant, and ${\mathsf k}_B T^{(X)} (a_0) = \biggl [\sqrt{\frac{2}{\pi}}~
\frac {(2\pi)^2}{m_X^{5/2}} \rho_X (a_0)\biggr ]^{2/3}$.

The distribution of $A$ particles which are produced during the decay of $X$ is also 
non-thermal. They are expected to be relativistic at early times if $m_A \ll m_X$ and 
non-relativistic at late times. Assuming that soon after they production they behave as a 
perfect fluid, the conservation equation gives their density, and thereby we find an 
effective temperature for them:
\be
{\mathsf k}_B T_A \approx \biggl [\frac{\pi^2}{3 \zeta (4)}{\mathcal M}_A \rho_X (\bar{x}) 
\frac{a_0}{a(t)} \biggl (\frac{a}{a_0} [\frac{\sqrt{\pi}}{2} - \exp ({-\frac{t-t_0}
{\tau}})] - \frac{\sqrt{\pi}}{2} + 1 \biggr )\biggr ]^{\frac{1}{4}} \label{atemprel}
\ee
where we have assumed a radiation dominated universe. When $A$ particles 
become non-relativistic their effective temperature can be estimated as:
\be
{\mathsf k}_B T_A \approx \biggl [\sqrt{\frac{2}{\pi}}~
\frac {(2\pi)^2}{m_A^{5/2}}{\mathcal M}_A \rho_X (\bar{x}) (1- 
\exp ({-\frac{t-t_0}{\tau}})) \biggr ]^{\frac{2}{3}} \label{atempnonrel}
\ee
Note that in these approximations are obtained with the assumption that $\tau \gg \tau_U$, 
i.e. a very slowly decaying $X$. When $\tau \ll \tau_U$ they should valid approximations 
when $t \ll \tau$. We emphasis again that the thermal distributions and temperatures 
calculated here are simple prescriptions when we can not solve all the evolution equations 
consistently.

The case of $\Phi$ particles is somehow different because some of them join the condensate. 
Therefore, the effective temperature of non-condensate component depends on $\varphi$. 
Nonetheless, if we neglect their small coupling to other species, the total energy in the 
two components must be equal to the energy transformed to $\Phi$ during the decay of $X$. 
Their effective temperature have the same form as (\ref{atemprel}) but with a branching 
factor that presents the fraction of energy transformed to non-condensated particles:
\be
{\mathcal M}_\Phi \rightarrow {\mathcal M}_\Phi (1 - \frac{2 \tau \dot{\varphi} 
\ddot{\varphi} + V'(\varphi)}{\rho_X}) \label{tempfrac}
\ee
where we have assumed that during $X$ particles decay half of the energy is transferred 
to $\Phi$ and the other half to $A$. We have also neglected the spacial fluctuations of 
the energy density of $X$ and $\varphi$.

When we calculate the propagators of $\phi$ we should take into account the contribution 
of all $\Phi$ particles in the wave function $\Psi$, including the condensate. Therefore:
\be
|\Psi^{(\Phi)}|^2 \approx f^{(\Phi)} (p, \bar{x}) + f^{(\varphi)}(\bar{x}) 
\label{psiphi}
\ee
where $f^{(\varphi)}$ is the contribution of the condensate. Note that the separation of 
two components in (\ref{psiphi}) is an approximation and ignores the quantum interference 
between them. It is valid if the self-interaction of $\Phi$ is weak and the 
non-condensate component decohere quickly.

We ignore a general description for the wave function of the condensate component. 
Nevertheless special cases can be found, see e.g. ~\cite{condwave} and Appendix 
\ref{app:b}. These states are special 
cases in which the addition of more particles does not change the state. In another word, 
they have a zero chemical potential. However, a condensate can be formed only in an 
interacting system, and in general the chemical potential is not zero. Using the 
description of the condensate wave function suggested in~\cite{condwave}~\footnote{it is evident that expression (\ref{condwavef}) is inspired from Bose-Einstein condensates in condense 
matter in which bosons with similar quantum numbers share the same energy state. Although 
this state satisfies the definition of a condensate according to (\ref{classphi}), no 
constraint on energy states or else can be concluded from the latter definition. 
Therefore definition (\ref{condwavef}) is a special case.}:
\be
|\Psi_C\rangle \equiv e^{-|C|^2} e^{C a_0^{\dagger}} |0\rangle = e^{-|C|^2} \sum_{i=0}^\infty 
\frac {C^i(x)}{i!}(a_0^{\dagger})^i |0\rangle \label{condwavef}
\ee
It is easy to verify that this state satisfies the relation~\cite{condwave}:
\be
a_0 |\Psi_C\rangle = C |\Psi_C\rangle \label{condcond}
\ee
From decomposition of $\phi$ to creation and annihilation operators (\ref{canon}) we find:
\be
\chi (x) \equiv a \langle \Psi_C |\Phi|\Psi_C\rangle = 
C \um_0 (x) + C^* \um_0^* (x)\label{condexp}
\ee
Here we have adapted the original formula of~\cite{condwave} for a homogeneous FLRW 
cosmology. As $\chi$ is a real field the argument of $C$ is arbitrary and therefore 
we assume that $C$ is real:
\be
C = \frac{\um_0 (x) + \um_0^* (x)}{\chi (x)} \label{condc}
\ee 

Assuming that the wave function of the condensate can be factorized, it is 
clear that in the expressions (\ref{propfu}) and (\ref{proppa}) for the propagators, the 
condensate contributes only in terms in which at least some of the momentums are zero. 
Neglecting the interaction between particles, the wave function of these particles 
can be factorized and expressed as $f^{(\phi)}(p, x)$ (see e.g.~\cite{thermalprop}). The 
advanced and retarded propagators of $\phi$ can be written as\footnote{Note that for the 
condensate the momentum vector $k$ presents its Fourier transform. By contrast, in 
$f (k,x)$ which is a classical distribution, $k$ presents the momentum.}:
\bea
iG^{(\Upsilon)>} (x,y) & = & \frac{1}{(2\pi)^3} \int d^3 p 
\biggl [f^{(\phi)}(p, \bar{x})~\um_p^* (x)\um_p (y) + (1 + f^{(\phi)}(p, 
\bar{x}))~\um_p (x)\um_p^* (y) \biggr ] + \nonumber \\
&& |C (\bar{x})|^2 \um_0^* (x)\um_0 (y) + (|C (\bar{x})|^2 + 1) \um_0 (x)\um_0^* (y) 
\label{propfuphi} \\
iG^{(\Upsilon)<} (x,y) & = & \frac{1}{(2\pi)^3} \int d^3 p 
\biggl [f^{(\phi)}(p, \bar{x})~\um_p (x)\um^*_p (y) + (1 + f^{(\phi)}
(p, \bar{x}))~\um_p^* (x)\um_p (y) \biggr ] + \nonumber \\
&& |C (\bar{x})|^2 \um_0 (x)\um_0^* (y) + 
(|C (\bar{x})|^2 + 1) \um_0^* (x)\um_0 (y) \label{proppaphi} 
\eea
where $E^2 = P^2+ m^2$. For fields $X$ and $A$ which do not have a condensate \ref{propfuphi} 
and \ref{proppaphi} can be use with $C = 0$. 

In Sec. \ref{sec:classevol} we used the Fourier Transform (FT) of the propagators in the 
quantum correction of the condensate evolution. To obtain the transformation of 
propagators in (\ref{propfuphi}) and (\ref{proppaphi}) we write them as the following 
(for $C = 0$):
\be
iG^{>} (x,y) = \frac{1}{(2\pi)^3} \int d^3 p \biggl [f(p, \bar{x})~
\um_p^* (\eta)\um_p (\eta') e^{i\vec{p}.\vec{X}} + (1 + f(p, \bar{x}))~\um_p (\eta)
\um_p^* (\eta') e^{-i\vec{p}.\vec{X}}\biggr ]
\ee
It is clear that $G^{>} (x,y)$ depends on both $X$ and $\bar{x}$, in contrast to Minkovski 
space-time where it dependens only on $X$. Nonetheless, it can be factorized to terms that 
depend only on one or the other coordinate. Therefore, the FT with respect to $X$ is 
defined as:
\bea
G^{>} (\bar{x},X) & = & \frac{1}{(2\pi)^3} \int d^3 p G^{>}_p (\bar{x},\eta,\eta') 
e^{-i\vec{p}.\vec{X}} \label{prodxxbar} \\
G^{>}_p (\bar{x},\eta,\eta') & \equiv & f(-p, \bar{x})~\um_{-p}^* (\eta)
\um_{-p} (\eta') + (1 + f(p, \bar{x}))~\um_{p} (\eta)\um_{p}^* (\eta') \label{progfourier}
\eea
We remind that propagators are used to determine the closed time path integrals and thereby 
the expectation values in the evolution equation of the condensate. This equation is a 
partial differential equation, and we need to take its FT with respect to one of the 
variables $x$ to solve it. If $G (x,y)$ had a translation symmetry, the FT with respect to 
$X$ variable was equal to the FT with respect to $x$ or $y$ up to a sign. But in an 
expanding non-empty universe there is no translation symmetry. Therefore, strictly speaking 
one can not use (\ref{progfourier}) in place of FT with respect to $x$. But, it is straight 
forward to show that the FT with respect to $x$ mixes the modes in the quantum correction 
terms and makes any analytical solution of the condensate evolution equation very difficult. 
On the other hand, the fact that $G^{>}_p (\bar{x},\eta,\eta')$ factorizes to components 
that depend only on one of the variables $\bar{x}$ and $X$ suggests that as an approximation 
we can identify (\ref{progfourier}) with the FT with respect to $x$, and treat $\bar{x}$ 
as an independent variable. Because the latter appears only in the energy distribution of 
particles, and $\um (x) \um^* (y)$ and its conjugate have translation symmetry this should 
be a good approximation. This allows to treat $G^{>}_p (\bar{x},\eta,\eta')$ like 
propagators in a Minkovski space-time. In Sec. \ref{sec:classevol} we use 
$G^{>}_p (\bar{x},\eta,\eta')$ as the FT of propagators during determination of the closed 
time path integrals.   

\section{Generalized multi-condensate state} \label{app:b}
Equation (\ref{condwavef}) which satisfies the definition (\ref{classphi}) for a condensate 
(classical field) can be generalized in the following manner: Consider a system with a large 
number of scalar particles of the same type. The only discriminating observable is their 
momentum. The distribution of momentum is discrete if the system is put in a finite volume. 
Such setup can contain sub-systems similar to (\ref{condwavef}) consisting of particles 
with momentum $\vec{k}$:
\be
|\Psi_k\rangle \equiv A_k e^{C_k a_k^{\dagger}} |0\rangle = A_k \sum_{i=0}^N 
\frac {C_k^i}{i!}(a_k^{\dagger})^i |0\rangle \label{condwavek}
\ee
where $A_k$ is a normalization constant. It is easy to verify that this state satisfies 
the relation:
\be
a_k |\Psi_k\rangle_N = C_k |\Psi_k\rangle_{(N-1)} \label{condcondk}
\ee
Therefore if $N \rightarrow \infty$, the identity (\ref{condcondk}) becomes similar to 
(\ref{condcond}) and the expectation value of the scalar field on this state is non-zero. 
However, if $|k| \neq 0$, the total energy of the system becomes infinite unless  
$k \rightarrow 0$ and $\Delta k \rightarrow 0$. Both here and the special condensate 
explained in Appendix \ref{app:a} the mass of condensate is assumed to very small. In most 
realistic physical systems with a condensate, these conditions do exist. Therefore, we 
define a multi-condensate or generalized condensate state as a state in which every particle 
belongs to a sub-state of the form (\ref{condwavek}): 
\bea
&& |\Psi_{GC}\rangle \equiv \sum_k A_k e^{C_k a_k^{\dagger}} |0\rangle = \sum_k A_k 
\sum_{i=0}^{N \rightarrow \infty} \frac {C_k^i}{i!}(a_k^{\dagger})^i |0\rangle 
\label{condwaveg} \\
&& \chi (x,\eta) \equiv a \langle \Psi_{GC}|\Phi|\Psi_{GC}\rangle = \sum_k  
C_k \um_k (x) + C^*_k \um_k^* (x) \label{condexpg}
\eea
The state $|\Psi_{GC}\rangle$ satisfies the equality(\ref{condcondk}). The coefficients $C_k$ 
determine the relative amplitudes of the single-particle states with different momentum. 
Using (\ref{condexpg}), the evolution equation of $\chi$ determines how $C_k$'s evolve.

The extension of propagators (\ref{propfuphi}) and (\ref{proppaphi}) to this state is trivial: 
\bea
iG^{\Upsilon >} (x,y) & = & \sum_k \biggl [f^{(\phi)}(k, \bar{x}) \um_k^* (x)\um_k 
(y) + (1 + f^{(\phi)}(k, \bar{x}))~\um_k (x)\um_k^* (y) + \nonumber \\ 
&& |C_k (\bar{x})|^2~\um_k^* (x)\um_k (y) + 
(1 + |C_k (\bar{x})|^2)~\um_k (x)\um_k^* (y) \biggr ] \label{propfuphigen} \\
iG^{\Upsilon<} (x,y) & = & \sum_k \biggl [f^{(\phi)}(k, \bar{x})~\um_k (x)\um^*_k (y) + 
(1 + f^{(\phi)}(k,\bar{x})~\um_k^* (x)\um_k (y) + \nonumber \\ 
&& |C_k|^2~\um_k (x)\um^*_k (y) + (1 + |C_k|^2)~\um_k^* (x)\um_k (y)
\biggr ] \label{proppaphigen} 
\eea
A simple example of such system can be the condensation of scalar particles in a potential 
well. Because only discrete energy levels are allowed, at equilibrium there can be a 
superposition of condensates with an effective mass (momentum) difference of $\Delta m_{eff} = 
n/L$ where $L$ is the size of the well.

In the context of the model explained here the existence of a condensate in which particles 
have different energies is important because this means that $\Phi$ particles do not need to 
lose completely their momentum to join the condensate. Such a state can potentially have 
applications in condense matter too, because in some sense it has simultaneously the 
properties of bosonic systems - condensation - and fermionic systems - a spectrum of energy 
levels. An example in the bulk excitation of a Bose-Einstein Condensate 
(BEC)-Bardeen-Cooper-Schreffer super fluidity~\cite{bec}.

\section{Propagators in matter dominated epoch} 
\label{app:c}
In the matter dominated epoch the relation between comoving and conformal time is:
\bea
 \eta = \int \frac{dt}{a} = \eta_0 \biggl( \frac{t}{t_0}\biggr )^{\frac{1}{3}} 
&,& \eta_0 \equiv \frac{3t_0}{a_0} \label{etamatter} \\
 \frac{a}{a_0} = \biggl( \frac{t}{t_0}\biggr )^{\frac{2}{3}} = 
\biggl( \frac{\eta}{\eta_0}\biggr )^2 &,& \frac{a''}{a} = \frac{2}{\eta^2} 
\label{amatter}
\eea
By applying (\ref{amatter}) to the Green's function equation (\ref{propagx}) the field 
equation for the modes gets the following form:
\be
\um_k'' + (k^2 + \frac{m^2 a_0^2\eta^4}{\eta_0^4} - \frac{2}{\eta^2}) \um_k = 0 
\label{progmatter}
\ee
where $\um_k$ presents the solution for one of the field $X$ or $A$. For two special cases of 
$m = 0$ and $k^2 = 0$ this equation has exact analytical solutions:
\be
\um (\eta) = \begin{cases} \sqrt{\frac{\eta}{\eta_0}} J_{\pm \frac{1}{2}} (\beta' 
\frac{\eta^3}{\eta_0^3}) \quad , \quad 
\beta' \equiv \frac{a_0\eta_0 m}{3} = \frac{2m}{3H_0} & \text{For $k^2 = 0$}  \\
\sqrt{\frac{\eta}{\eta_0}} J_{\pm \frac{3}{2}} (k \frac{\eta}{\eta_0}) & 
\text{For $m = 0$} \end{cases} \label {solkmmatter}
\ee
During matter domination epoch the masses of $A$ and $X$ particles are considered to be much 
larger than their kinetic energy and the Hubble constant. Moreover, in a cosmological context 
only large scales with $k \ll m$ are under scrutiny. Therefore, we use $k=0$ solution as the 
zero-order approximation and use the WKB-like techniques to find an approximation for 
$k \neq 0$ case. Only the argument of the Bessel function in (\ref{solkmmatter}) is 
mass-dependent. Thus, we replace it with a WKB-like integral:
\bea
\frac{a_0\eta_0 m}{3} \frac{\eta^3}{\eta_0^3} & \rightarrow & a_0\eta_0 \int d(\frac{\eta}
{\eta_0}) \frac{\eta^2}{\eta_0^2}\sqrt{m^2 + \frac{k^2}{a^2}} \label{wkbu} \\
\um_k (\eta) & \approx & \sqrt{\frac{\eta}{\eta_0}} J_{\pm \frac{1}{2}} \biggl [\beta' 
\frac{\eta^3}{\eta_0^3} (1- \frac{3k^2 \eta_0}{2m^2 \eta}) \biggr ] \label {solmmatterbess}
\eea
The Bessel functions $J_{\pm \frac{1}{2}}$ have analytical expressions:
\be
J_{\frac{1}{2}} (x) = \sqrt{\frac{2}{\pi x}} \sin x , \quad 
J_{-\frac{1}{2}} (x) = \sqrt{\frac{2}{\pi x}} \cos x \label {besselonehalf}
\ee
After including (\ref{besselonehalf}) to (\ref{solmmatterbess}) we obtain the following 
approximate expression for the solution of (\ref{progmatter}):
\be
\um_k (\eta) \approx \sqrt{\frac{2}{\pi \beta'}}~\frac{\eta_0}{\eta}~(1 - 
\frac{3k^2\eta_0}{2m^2\eta})^{-\frac{1}{2}} \biggl [c_k \sin \biggl (\frac{\beta' \eta^3}
{\eta_0^3}(1- \frac{3k^2\eta_0}{2m^2\eta})\biggr ) + d_k \cos \biggl (\frac{\beta' \eta^3}
{\eta_0^3}(1-\frac{3k^2\eta_0}{2m^2\eta}) \biggr ) \biggr ] \label {homosolmatter}
\ee
Boundary and initial conditions explained in Sec. \ref{sec:homosol} are also applied to 
(\ref{homosolmatter}).

In the case of the field $\phi$, the propagator (\ref {propup}) includes 
also an additional mass term that depends on the condensate field $\varphi$, and therefore 
on $\eta$. The effect of this term can be included in the same way as $k^2$ by 
replacing $k^2\eta_0/m_\Phi^2\eta$ terms with $k^2\eta_0/m_\Phi^2\eta + \lambda (n-1) 
\varphi^{n-2}/ m_\Phi^2$ in (\ref{homosolmatter}). If at late times $\varphi \rightarrow 
const$ - expected for a dark energy field - the contribution of this term does not vanish, 
in contrast to $k$-dependent terms. Thus, it can be added to the mass of the field 
and the definition of $\beta'$. This dynamical mass has the very important role of feedback 
on the growth of the condensate, see Sec. \ref{sec:matdom} for details.

\section{Propagators and condensate evolution in a fluctuating 
background} \label{app:d}
Using the metric (\ref{metric}), the Green's function equation (\ref{propagphi}) for the 
propagator $G_F (x,y)$ can be written as (we drop the field index for simplicity):
\bea
&& (1-2\psi)^{\frac{1}{2}} G''_F - \frac{\psi' (G'_F - G_F a'/a)}
{(1-2\psi)^{\frac{1}{2}}} - \delta^{ij} \biggl [\frac{\partial_i \psi \partial_j G_F}
{(1+2\psi)^{\frac{1}{2}}} + (1+2\psi)^{\frac{1}{2}} \partial_i \partial_j G_F \biggr ] + 
\nonumber \\
&& \hspace{2cm} \biggl [(1-4\psi^2)^{\frac{1}{2}}(a^2 m^2_\Phi + (n-1) \lambda a^{4-n}
\chi^{n-2}) - (1-2\psi)^{\frac{1}{2}} \frac{a''}{a}\biggr ] G_F = 
-i \frac{\delta^4 (x-y)}{a} \label{progphifluc}
\eea
Because (\ref{progphifluc}) is a linear differential equation and the metric 
fluctuation $\psi$ is small $G_F$ can be decomposed to:
\be
G_F (x,y) = G_F^h (x,y) + \Delta G_F (x,y) \quad, \quad \biggl |\frac{\Delta 
G_F (x,y)}{G_F^h (x,y)} \biggr | \ll 1 \label{gfdecomp} 
\ee
where $G_F^h (x,y)$ is the propagator in a homogeneous background i.e. the 
solution of equation (\ref{propup}). After inserting this definition in 
(\ref{progphifluc}), we find the following equation for $\Delta G_F$:
\bea
&& \Delta G''_F - \delta^{ij} \partial_i \partial_j \Delta G_F + 
(a^2 m^2_\Phi + (n-1) \lambda a^{4-n}\chi^{n-2} - \frac{a''}{a}) \Delta G''_F = 
-i \frac{\psi}{a} \delta^4 (x-y) + \zeta (x-y) \label{progphiflucdelt} \\
&& \zeta (x-y) \equiv \psi'(G^{'h}_F - G_F^h a'/a) + \delta^{ij}(\partial_i \psi 
\partial_j G_F^h + 2 \psi \partial_i \partial_j G_F^h) - \psi \biggl [a^2 m^2_\Phi + 
(n-1) \lambda a^{4-n}\chi^{n-2} \biggr ] G_F^h \nonumber \\
&& \label{theta}
\eea
The left hand side of (\ref{progphiflucdelt}) is similar to (\ref{propup}), thus both 
equations have the same homogeneous solution. If we neglect $\zeta$ term, the only difference 
between (\ref{progphiflucdelt}) and (\ref{propup}) is the $\psi (y)$ factor in front of the 
delta-function in the right hand side of (\ref {progphiflucdelt}). Therefore, the solutions 
of these equations are the same up to a $\psi (y)$ factor. Because without $\zeta$ term the 
equation (\ref{progphiflucdelt}) has the form of a Green's function, its solution in presence  
of a non-homogeneous term like $\zeta$ is:
\be
\Delta G_F (x,y) = \psi (y) G_F^h (x,y) + \psi G_F^h \otimes 
\zeta (x,y) \label{deltagsol}
\ee
The second term in (\ref{deltagsol}) is of second order, and thus at first order in 
fluctuations:
\be
G_F (x,y) = (1+\psi) G_F^h (x,y) \label{propfluc}
\ee
Under the approximations considered in Sec. \ref{sec:classevol} when we calculate the closed 
time path integrals, the factors $(1+\psi)$ of propagators can be included into 
$A(k,\bar{x})$ in (\ref{wksolint}) for radiation domination, and into $C (k,\bar{x})$ in 
(\ref{vmsol}) for matter domination. As explained in Sec. \ref{sec:classevol}, in this 
approximation the coordinate variable $x$ is identified with $\bar{x}$.

The evolution equation for the classical component of $\Phi$ must be also written for the 
background metric (\ref{metric}):
\bea
&& {\chi}'' - \frac {\psi' (\chi' - \chi a'/a)}{(1 - 2 \psi)} - \delta_{ij}
\biggl [\frac{\partial_i \psi \partial_j \chi}{(1-4\psi^2)^{\frac{1}{2}}} + 
(\frac {1+2\psi}{1-2\psi})^{\frac{1}{2}} \partial_i\partial_j \chi \biggr ] + \nonumber \\
&& \hspace{1cm} \biggl [a^2 m_{\Phi}^2 
(1-4\psi^2)^{\frac{1}{2}} - (1-2\psi)^{\frac{1}{2}} \frac{a''}{a} \biggr ] + 
\frac{\lambda a^{4-n}}{n}\sum_{i=0}^{n-1} (i+1) \binom{n}{i+1} {\chi}^i \langle 
{\Upsilon}^{n-i-1}\rangle - \omega (x) = 0 \nonumber \\
&& \label {chiperturb} \\
&& \omega (x) = \begin {cases} a \mg \langle \cx\ca \rangle & 
\text{For (\ref{decaymode})-a} \\ 
\mg \langle \cx\ca^2 \rangle = 0 & \text{For (\ref{decaymode})-b} \\
2 (\mg \chi \langle \cx\ca \rangle + \mg \langle \Upsilon \cx\ca \rangle) & 
\text{For (\ref{decaymode})-c} \end{cases} \label{omegadef}
\eea
In contrast to (\ref{progphifluc}), for $n > 2$ this equation is non-linear and a 
decomposition similar to (\ref{gfdecomp}) is not allowed. In the present work the aim of 
solving equation analytically obliged us in many places to neglect the coordinate dependence 
and non-linear terms. Therefore, in this work we neglect metric fluctuations in 
(\ref{chiperturb}). This leads to equations (\ref{evola})-(\ref{evola}).
\end{appendix}
\end{fmffile}

\end{document}